\documentclass[a4paper]{report}
\usepackage[utf8]{inputenc}
\usepackage[T1]{fontenc}
\usepackage{RJournal}
\usepackage{amsmath,amssymb,array}
\usepackage{booktabs}
\usepackage{graphicx}
\usepackage{xcolor}

\begin{document}

\sectionhead{Contributed research article}
\volume{XX}
\volnumber{YY}
\year{20ZZ}
\month{AAAA}

\begin{article}

\title{\pkg{binspp}: An R Package for Bayesian Inference for Neyman-Scott Point Processes with Complex Inhomogeneity Structure}
\author{by Jiří Dvořák, Radim Remeš, Ladislav Beránek and Tomáš Mrkvička}

\maketitle

\abstract{
The Neyman-Scott point process is a widely used point process model which is easily interpretable and easily extendable to include various types of inhomogeneity. The inference for such complex models is then complicated and fast methods, such as minimum contrast method or composite likelihood approach do not provide accurate estimates or fail completely. Therefore, we introduce Bayesian MCMC approach for the inference of Neymann-Scott point process models with inhomogeneity in any or all of the following model components: process of cluster centers, mean number of points in a cluster, spread of the clusters. We also extend the Neyman-Scott point process to the case of overdispersed or underdispersed cluster sizes and provide a Bayesian MCMC algorithm for its inference. The R package \pkg{binspp} provides these estimation methods in an easy to handle implementation, with detailed graphical output including traceplots for all model parameters and further diagnostic plots. All inhomogeneities are modelled by spatial covariates and the Bayesian inference for the corresponding regression parameters is provided.
}


\section{Introduction}
The Neyman-Scott point process \citep{NeymanScott1958} is a cluster point process model widely used in biology, astronomy, forestry, medicine, etc. Its main advantages are the straightforward interpretation of the model parameters and the closed form of moment properties, at least for the stationary version of the model.

The model can be constructed in two stages and it is often called a doubly stochastic process: first, cluster centers are randomly sampled from a given Poisson point process; second, conditionally on the positions of cluster centers, random number of offspring points are randomly and independently placed around each cluster center. Thus, the stationary Neyman-Scott point process model is specified by the intensity of the Poisson process of cluster centers, the distribution of the number of points per cluster (hereafter called cluster size), and the distribution of the relative displacement of the offspring points around their respective cluster centers (hereafter called cluster spread).

In the most popular type of Neyman-Scott point process, called the (modified) Thomas process \citep{Thomas1949}, the cluster size is assumed to follow a Poisson distribution and the cluster spread is governed by a radially symmetric Gaussian distribution. The stationary Thomas process is then described by the following parameters: the intensity $\kappa$ of the process of cluster centers, the mean number of points in a cluster $\alpha$ and the standard deviation $\omega$ of the Gaussian distribution determining cluster spread.

A natural way of introducing inhomogeneity into the Neyman-Scott process model is allowing some of the model components (intensity of cluster centers $\kappa$, cluster size $\alpha$, cluster spread $\omega$) depend on a set of spatial covariates. Their significance then needs to be assessed. Several models of this type have been studied in the literature, as discussed in the following paragraphs.

The Neyman-Scott point process with inhomogeneous cluster centers, with the distribution of the clusters being the same in terms of size and spread, allows for varying the number of clusters in the space. The inference for this process was investigated in \citet{MMK2014} and it was found that the Bayesian MCMC estimation procedure is more precise than composite likelihood or minimum contrast method. This model is not second-order inhomogeneity reweighted stationary (SOIRS) \citep{BMW2000}, but if instead the mean number of points in a cluster $\alpha$ is inhomogeneous, the resulting process is very close to SOIRS. The inference for SOIRS Neyman-Scott process (stationary Neyman-Scott point process thinned by a spatially varying function) can be performed by a two step method based on minimum contrast or composite likelihood \citep{WG2009}, as implemented in the R package \CRANpkg{spatstat} \citep{spatstat}. 

The inhomogeneity can be also introduced in the cluster spread $\omega$, then the process will be locally-scaled Neyman-Scott point process \citep{HJLN2003}. It is also possible to introduce inhomogeneity simultaneously in  $\alpha$ and $\omega$, then the process can be called a  Neyman-Scott point process with growing clusters \citep{M2014}.

For the models with homogeneous population of cluster centers but inhomogeneity in the cluster properties we talk about cluster inhomogeneity. On the other hand, for the Neyman-Scott point process with inhomogeneous cluster centers we talk about inhomogeneity of centers. Combining the cluster inhomogeneity and inhomogeneity of centers leads to the notion of doubly inhomogeneous Neyman-Scott point process. The Bayesian MCMC inference for such a process was studied in \cite{MS2017}. Other kinds of inference were found to be useless for such complex model.

Therefore, we build up the R package \pkg{binspp} which contains the Bayesian MCMC estimation procedure for the most general model with inhomogeneity in all three model components. This model contains all the previous models as special cases, including the stationary one. The radially symmetric Gaussian distribution is assumed to determine the cluster spread, which does not limit practical applicability of the models. We do not include SOIRS Neyman-Scott point process due to the different construction of the model and also due to availability of moment-based estimation methods. However, practically speaking the Neyman-Scott point process with inhomogeneous cluster size $\alpha$ can be used as an approximation to the SOIRS Neyman-Scott point process model.

The most important statistical problem here is to assess the dependence of the data on the given set of covariates. Our package handles the spatial covariates influencing any or all of the model components: the intensity of the cluster centers $\kappa$, the cluster size $\alpha$ and the cluster spread $\omega$. The Bayesian MCMC procedure is time consuming, but its great benefit is that the significance of all covariates is provided from the estimated posterior distributions in a natural way. For example, in case of the inhomogeneity of cluster centers if a faster estimation method is used, it is necessary to perform parametric bootstrap in order to obtain the significance of the covariates. This is as much time consuming as the Bayesian MCMC procedure \citep{MMK2014}. Only in the case of SOIRS Neyman-Scott processes it is possible to use the fast estimation method and the significance of covariates can be assessed using the asymptotic normality result obtained in \citet{WG2009}.

As mentioned above, the minimum contrast approach and the composite likelihood approach are available for SOIRS Neyman-Scott point process. They are also described in \citet{MMK2014} for the Neyman-Scott point process with inhomogeneous cluster centers. To the our best knowledge, they are not available for the other kinds of inhomogeneity discussed above. 

In order to model inhomogeneous clustered point patterns log-Gauussian Cox process (LGCP) models are often used \citep{MSW1998}. The inference for SOIRS LGCP is well developed with integrated nested Laplace approximation (INLA) \citep{RMCh2009} available through the package \CRANpkg{inlabru} \citep{inlabru}, with Bayesian MCMC inference available through the package \CRANpkg{lgcp} \citep{lgcp} or with the moment methods available through the package \pkg{spatstat}. Nevertheless, none of these packages allows for inference for LGCP with more complex types of inhomogeneity. An attempt in this direction has been made in \cite{Dvoraketal2019}.

With these considerations in mind, we have implemented the core functions of the \pkg{binspp} package using \pkg{Rcpp} to speed-up the computation. As a result, short runs of the chain, useful for tuning up the hyperparameters of the prior distributions, are finished within minutes on a regular laptop. Long runs, used for the actual estimation, may be finished within a few hours, see the detailed example in Section~\ref{subsec:complex_example}. This makes our implementation easily applicable in practice, without the need to worry about the computational demands.

The inference for Neyman-Scott point processes is usually performed with the assumption of Poisson distribution of of the number of points in a cluster. The same is assumed in all models discussed above, but the \pkg{binspp} package contains also the Bayesian MCMC estimation procedure for homogeneous generalised Neyman-Scott process. The method was described in \cite{AM2020}. This model uses the generalised Poisson distribution (GPD) as a distribution of the number of points in a cluster. The GPD allows for modelling of under- or over-dispersion. This allows for more flexible modelling of the distribution of the number of points in a cluster.

The package specifically considers the following models. The inhomogeneous Thomas point process allowing for modelling of cluster centers, cluster spread and cluster sizes through covariates. The covariates are incorporated in every model component via exponential model. Such a general process was not presented in the literature, even \cite{MS2017} presents only a special case of this general model. Furthermore, the generalised Neyman-Scott point process which allows for modelling underdispersed or overdispersed cluster sizes is considered.

This paper is organized as follows. First, in Section~\ref{models} we describe the models considered here. Then we briefly describe the algorithms in Section~\ref{MCMC}. In Section~\ref{examples} we show the use of our package for different types of models, with a detailed example in Section~\ref{subsec:complex_example} considering a real dataset of infected oak trees. This is a part of a larger dataset studied in \citet{FFCGA2019}. The model for the observed point pattern includes inhomogeneity in all three components of the model and for illustration we provide the outputs of a long run of the MCMC chain. We also show in Section~\ref{subsec:generalised} the use of the estimation procedure for homogeneous generalised Neyman-Scott point process and the possibility of detecting over- or under-dispersion of cluster sizes. Section~\ref{sec:discussion} is left for discussion.

\section{Models} \label{models}

Let us first describe the model in its full generality, i.e. the doubly inhomogeneous Neyman-Scott point process. The process of cluster centers $C$ follows an inhomogeneous Poisson point process
with intensity function $\kappa f (\beta, u), \ u\in \mathbb{R}^2$, where $\kappa > 0$ 
and $\beta \in \mathbb{R}^k$ are parameters. The clusters $X_c, c \in C$, 
are independently attached to every cluster center $c$. 
The Neyman-Scott point process is the superposition of the clusters 
$X=\cup_{c\in C} X_c$, where $X_c$ 
are independent Poisson point processes with intensity function 
$\alpha (\mu,c)k(u-c, \omega (\nu,c)), u \in \mathbb{R}^2$,
which depends on $c$.
Here $\alpha(\mu, c)$ is the expected number of offspring points in the cluster corresponding to the parent point $c$ and $\mu \in \mathbb{R}^{l+1}$ is a 
parameter. Furthermore, $k(u, \omega(\nu, c))$ is the probability density function governing the relative displacement of the offspring points around the parent 
point $c$ (in this paper we assume $k$ is the density of the centered radially symmetric normal distribution with the standard deviation $\omega(\nu, c$)). The 
distribution of the number of points in the cluster with the cluster center $c$ is assumed to be Poisson for all inhomogeneous models, with the probabilities being 
denoted $p(n, \alpha(\mu, c))$.

The parametric functions $f$, $\alpha$ and $\omega$ are the key ingredients of the model which describe the dependence on the spatial covariates. These functions are assumed to take the following parametric form:
\begin{align*}
    f(\beta,u) & = \exp(\beta_1 z_1(u)+\ldots + \beta_k z_k(u)), \\
    \alpha(\mu,c) & = \exp(\beta^\alpha_0+\beta^\alpha_1 z^\alpha_1(c)+\ldots + \beta^\alpha_l z^\alpha_l(c)), \\
    \omega(\nu,c) & = \exp(\beta^\omega_0+\beta^\omega_1 z^\omega_1(c)+\ldots + \beta^\omega_m z^\omega_m(c)).
\end{align*}

Here $z_1, \ldots , z_k$ are the spatial covariates influencing the population of cluster centers, $z_1^\alpha, \ldots , z_l^\alpha$ are the spatial covariates influencing the cluster size and $z_1^\omega, \ldots , z_m^\omega$ are the spatial covariates influencing the cluster spread. All the $\beta$s are real-valued regression parameters. Note that the model for $f$ does not contain the intercept since its role is taken by the parameter $\kappa$. We use this parametrization to be consistent with the earlier works describing the models and the corresponding Bayesian inference \citep{M2014,KM2016,MS2017}.

Specific choices of $k,l,m$ result in different special cases of the general model. For $k = 0, l = 0, m = 0$ the process is the stationary Neyman-Scott process. For $k > 0, l = 0, m = 0$ we obtain the inhomogeneous cluster centers. Similarly, $k = 0, l > 0, m = 0 $ results in inhomogeneous cluster sizes, while $k = 0, l = 0, m > 0$ leads to locally scaled process (inhomogeneous cluster spread). For $k = 0, l > 0, m > 0$ we obtain the process with growing clusters \citep{M2014} and finally for $k > 0, l > 0, m > 0$ we obtain the most general, doubly inhomogeneous process \citep{MS2017}.

Our package allows for all the possible choices of $k, l, m$. However, the user must be aware of the possible identifiability issues which occur if the same covariate is used for $z_i$ and $z_j^\alpha$ for some $i$ and $j$, i.e. if the same covariate influences both $f$ and $\alpha$. It is a property of the two-step estimation algorithm described in the next section that the parameters $\beta_i$ and $\beta_j^\alpha$ cannot be estimated correctly in this case. For example, if $\beta_i = 0$ and $\beta_j^\alpha \neq 0$ then the first step of the algorithm will estimate $\hat\beta_i$ to be approximately $\beta_j^\alpha$.

Our \pkg{binspp} package allows also for the estimation of the homogeneous generalised Neyman-Scott point process, where the distribution of the cluster sizes follows the generalised Poisson distribution, which is a popular model for count data \citep{WF1997}. The probability mass function of GPD is
\begin{align*}
p(n|\lambda, \theta) = \left\{\begin{array}{l} \frac{1}{n!}\theta(\theta + \lambda n)^{n-1}e^{-\theta-\lambda n}, \quad  n = 0, 1, 2, \ldots, \\ 
0, \quad \quad  \text{if } n > m \text{ when } \lambda <0,
\end{array} \right.
\end{align*}
with the expectation $\alpha =  \frac{\theta}{1-\lambda}, \label{eq:E} $ and variance $\frac{\theta}{(1-\lambda)^3}. \label{eq:V}$ The parameter $\lambda \in [-1,1]$ in GPD models the over- or under-dispersion, $\lambda=0$ corresponds to the Poisson case, $\lambda>0$ to the over-dispersed and $\lambda<0$ to the under-dispersed case.

\section{Inference} \label{MCMC}
The inference for the doubly inhomogeneous Neyman-Scott point process is performed in two steps, similar to the approach of \cite{WG2009}.

In the first step the parameters $\overline{\beta}=(\log (\lambda)
,\beta_1, \ldots, \beta_k)$ of the intensity function are estimated, based on the assumption that $X$ is the
Poisson process with the intensity function
\begin{align} \label{aprox} \overline{f}_{\overline{\beta}}
(u) =  \exp (\overline z(u) \overline \beta^T), \  u \in \mathbb R^2,
\end{align}
    where $\overline z(u) = (1, z_1(u), \ldots , z_k(u))$.
This assumption is intuitively justified if the range of interaction among the points is small compared to the range of changes in the spatial covariates and if $z_i$ is different form $z_j^\alpha$ for all combinations of $i$ and $j$.
Specifically, we maximize the log-likelihood of the assumed inhomogeneous Poisson process:
\begin{align}
    l(\overline{\beta})=\sum_{x\in X \cap W}
    \overline z(x)\overline{\beta}^T - \int_W \exp (\overline z(u)\overline{\beta}^T)
    \, \mathrm{d}u
\end{align}
Here $W$ is the observation window.

The second step consists of estimation of the interaction parameters $\mu$
and $\nu$, conditionally on the estimate of $\overline{\beta}$. 

Bayesian estimation for the homogeneous Neyman-Scott point processes was carried out with an MCMC algorithm e.g. in \cite{GT2012, MW2007, M2014, KM2016}. In this approach, the cluster centers and the model parameters are updated in each step of the MCMC algorithm. After reaching the equilibrium, posterior distributions of the parameters can be estimated. The cluster centers are generally viewed as nuisance parameters.

Considering the inhomogeneous clusters, the MCMC algorithm proceeds in the same way as in the homogeneous case, except that the likelihood is influenced by the parameters connected with the cluster inhomogeneity. We remark here that the estimation algorithm was not presented in such generality earlier, even \cite{MS2017} considered only a special case of this model, without allowing for the general form with covariates. However, the generalisation is straightforward.

Let $C$ denotes the inhomogeneous Poisson point process of cluster centers with the intensity $\kappa f(\beta, u)$. 
Let $p(C|\kappa, \beta)$ denote the Poisson probability
density function of the point process $C$, conditionally on $\kappa$ and $\beta$,
with respect to the unit-rate homogeneous Poisson point process. Furthermore, let  $p(X|C, \beta, \kappa, \mu, \nu)$ denote the Poisson probability
density function of the point process $X$ under the knowledge of $C$, $\beta$, $\kappa$, $\mu$ and $\nu$. The joint posterior distribution of the process $C$ and the parameters is then
\begin{align}
  p(C, \kappa, \mu, \nu |X) \propto p(X|C, \beta, \kappa, \mu, \nu) p(C|\kappa, \beta)  p(\mu) p(\nu),
\end{align} 
where $p(\mu)$ and $p(\nu)$ denote the prior probability density functions for the respective parameters. No prior for $\kappa$ is required because  it is, in our estimation procedure including a modification similar to the one proposed by \citet{KM2016}, a deterministic function of $\mu$ and $\beta$. Indeed, the expected number $\mathbb E M$ of the observed points in the observation window $W$ is equal to  
\begin{align*}
  \kappa  \int_W \alpha(\mu, u) [\int_{\mathbb R^2} k(u-c,\omega (\nu, c))f(\beta, c) \, \mathrm{d}c ] \, \mathrm{d}u,
\end{align*}
which can be approximated by 
\begin{align*}
    \mathbb E M \approx \kappa  \int_W \alpha(\mu, u))f(\beta, u) \, \mathrm{d}u.
\end{align*}
Thus, in each iteration of the MCMC algorithm, $\kappa$ can be re-computed when $\mu$ is updated. 

Our MCMC algorithm consists of updating the process of cluster centers $C$ and updating the parameters $\mu$ and $\nu$. For updating $C$ we use the birth-death-move algorithm described in \citet{MollerWaagepetersen2004}. For updating $\mu$ and $\nu$ we use the Metropolis-Hastings algorithm. In order to obtain better mixing properties $\mu$ and $\nu$ is updated separately. Full details about this algorithm can be found in \cite{MS2017}.

The inference for the interaction parameters $\mu$ and $\nu$ is performed from the estimated posterior distributions which are obtained from the MCMC samples after appropriate burn-in. The inference about the first-order inhomogeneity parameters $\beta$ cannot be obtained from the first step where the Poisson distribution is assumed. Therefore, we base the inference about $\beta$ on the posterior distribution of the cluster centers obtained in the second step. In every step the significance of $z_1, \ldots, z_k$ with respect to the process of cluster centers $C$ is assessed using the Poisson distribution of $C$ assumed in this model. The median of respective $p$-values computed from the posterior distribution is taken to be the estimate of the $p$-value of the test of significance of the given covariate. We remark that it is also possible to perform the full Bayesian estimation by considering the inhomogeneity of cluster centers in the MCMC procedure. However, this approach was found to be less efficient than the two-step approach due to identifiability issues in the full likelihood.

Considering the generalised Neyman-Scott process, the MCMC algorithm consists of one extra step in addition to the traditional Metropolis-Hastings update of the model parameters and the birth-death-move update of the cluster centers. It is the update of the connections between the points and the cluster centers, since by assuming the non-Poisson distribution these connections take part in the likelihood of the process. Full details about this algorithm can be found in \cite{AM2020}.

\section{Examples} \label{examples}

In this section we provide a set of examples illustrating how different types of models can be fitted using the Bayesian MCMC approach implemented in the \pkg{binspp} package. The first few examples illustrate the use of the package, while the most complex example in Section~\ref{subsec:complex_example} describes the outputs of the algorithm in full detail. Model parametrization is described in Section~\ref{models}.

Below we assume that \texttt{X} is the observed point pattern in the \texttt{ppp} format used in the \pkg{spatstat} package. We further assume that \texttt{X} is observed through the observation window $W$, which is a union of aligned rectangles, aligned with the coordinate axes. \texttt{x\char`_left}, \texttt{x\char`_right}, \texttt{y\char`_bottom} and \texttt{y\char`_top} are vectors giving the coordinates of the extreme points of the rectangles whose union forms the observation window \texttt{W}. Furthermore, \texttt{W\char`_dil} is the dilated observation window used to accommodate cluster centers outside \texttt{W} to mitigate the edge effects. An easy way to obtain $W\char`_dil$ from the vectors \texttt{x\char`_left} to \texttt{y\char`_top} is shown in Section \ref{subsec:complex_example}. All covariates such as \texttt{cov1} are assumed to be pixel images (objects of type \texttt{im} from the \pkg{spatstat} package) defined over the \texttt{W\char`_dil} domain.

The list \texttt{control} contains important tuning constants such as the required number of iterations to be run (\texttt{NStep}), the length of the initial part of the chain to be discarded before computing estimates (\texttt{BurnIn}) or the sampling frequency used to reduce autocorrelations in the values used for computing the estimates (\texttt{SamplingFreq}). Also, hyperparameters for prior distributions for different parameters can be specified in this list, as illustrated in the detailed example in Section~\ref{subsec:complex_example}. Providing hyperparameter values guided by the knowledge of the problem at hand is highly recommended! However, some default values are used if the user does not provide them. 

All priors are normal distributions, priors for $\beta_i^\alpha$ and $\beta_i^\omega$, $i>0$, have expectation 0. We remark here that also the hyperparameters for $\beta_0^\alpha$ and $\beta_0^\omega$ must be given in $\log$ scale because all parameters, except of $\kappa$, are estimated in the exponential form.

\subsection{Homogeneous Thomas process}

Based on our experience, for this homogeneous model we recommend running at least $50\,000$ iterations, with burn-in of at least $25\,000$ steps.
\begin{example}
control = list(NStep=50000, BurnIn=25000, SamplingFreq=10)
\end{example}
The following commands provide equivalent ways of specifying that all three components of the model are homogeneous (no covariates are provided):
\begin{example}
Output = binspp(X, control, x_left, x_right, y_bottom, y_top, W_dil)
Output = binspp(X, control, x_left, x_right, y_bottom, y_top, W_dil, 
             z_beta=NULL, z_alpha=NULL, z_omega=NULL)
Output = binspp(X, control, x_left, x_right, y_bottom, y_top, W_dil, 
             z_beta=list(), z_alpha=list(), z_omega=list())
\end{example}

In this example default hyperparameter values were used. These can be retrieved in the following way:

\begin{example}
Output$priorParameters
\end{example}

The text outputs and graphical outputs are obtained as follows:
\begin{example}
print.outputs(Output)
plot.outputs(Output)
\end{example}

\subsection{Thomas process with inhomogeneous cluster centers}

In this example the population of parent points is inhomogeneous, with intensity function depending on a covariate. More covariates can be included in the model, provided they are given in the \texttt{z\char`_beta} list, see Section~\ref{subsec:complex_example} for an illustration.

For this model with simple inhomogeneity we recommend running at least $100\,000$ iterations, with burn-in of at least $50\,000$ steps. Note that the covariates in the list \texttt{z\char`_beta} must be named in order to the \texttt{ppm} function from the \pkg{spatstat} package run properly.

\begin{example}
control = list(NStep=100000, BurnIn=50000, SamplingFreq=10)
Output = binspp(X, control, x_left, x_right, y_bottom, y_top, W_dil, z_beta=list(Z1=cov1))
\end{example}

\subsection{Thomas process with inhomogeneous mean number of points in a cluster}

Now we assume that the mean number of points in a cluster depends on the position of the corresponding parent point $c$, with the function $\alpha(\mu,c)$ depending on a covariate. Again, more than one covariate may be used, provided they are given in the \texttt{z\char`_alpha} list.

For this model with simple inhomogeneity we recommend running at least $100\,000$ iterations, with burn-in of at least $50\,000$ steps.

\begin{example}
control = list(NStep=100000, BurnIn=50000, SamplingFreq=10)
Output = binspp(X, control, x_left, x_right, y_bottom, y_top, W_dil, z_alpha=list(cov1))
\end{example}

\subsection{Thomas process with inhomogeneous cluster spread}

In this case the spread of the clusters depends on the position of the corresponding parent point $c$, with the function $\omega(\nu,c)$ depending on a covariate. As before, more than one covariate may be given in the \texttt{z\char`_omega} list.

For this model with simple inhomogeneity we recommend running at least $100\,000$ iterations, with burn-in of at least $50\,000$ steps.

\begin{example}
control = list(NStep=100000, BurnIn=50000, SamplingFreq=10)
Output = binspp(X, control, x_left, x_right, y_bottom, y_top, W_dil, z_omega=list(cov1))
\end{example}

\subsection{Thomas process with complex inhomogeneities in all model components}\label{subsec:complex_example}

For this example we use the real dataset provided in the \pkg{binspp} package, see also \cite{FFCGA2019}:

\begin{example}
X = trees_N4
x_left = x_left_N4
x_right = x_right_N4
y_bottom = y_bottom_N4
y_top = y_top_N4
\end{example}

Several covariates accompany the observed point pattern. In the following lists we specify which covariates are assumed to influence which model components. Note that each model component depends on two covariates. Due to identifiability reasons the lists \texttt{z\char`_beta} and \texttt{z\char`_alpha} must be disjoint (no covariate can appear in both lists).

\begin{example}
z_beta = list(refor=cov.refor, slope=cov.slope)
z_alpha = list(tmi=cov.tmi, td=cov.td)
z_omega = list(slope=cov.slope, reserv=cov.reserv)
\end{example}

The observation window is given as the union of aligned rectangles, aligned with the coordinate axes:

\begin{example}
W = owin(c(x_left[1],x_right[1]),c(y_bottom[1],y_top[1]))
if(length(x_left)>=2){ 
  for(i in 2:length(x_left)){ 
     W2 = owin(c(x_left[i],x_right[i]),c(y_bottom[i],y_top[i])) 
     W=union.owin(W,W2)
   } 
 }
\end{example}

Dilated observation window:

\begin{example}
W_dil = dilation.owin(W,100)
\end{example}

The parameter 100 for the dilation specifies the width of the zone around $W$ where centers having offsprings in $W$ can occur. Since the Gaussian distribution for the offsprings is used, the width is infinite in theory, but for computational reasons we bound this region.

For this model with complex inhomogeneities we recommend running at least $250\,000$ iterations, with burn-in of at least $150\,000$ steps. Hyperparameter values for prior distributions can be specified as follows:

\begin{example}
control = list(NStep=250000, BurnIn=150000, SamplingFreq=10, Prior_alpha_mean=3, 
              Prior_alpha_SD=2, Prior_omega_mean=5.5, Prior_omega_SD=5, 
              Prior_alphavec_SD=c(4.25,0.012), Prior_omegavec_SD=c(0.18,0.009))
\end{example}

The following commands perform the MCMC estimation. With the required 250\,000 steps of the algorithm the computation takes approx. 2.5 hours on a regular laptop due to the implementation of the core functions using the \pkg{Rcpp} package.

\begin{example}
set.seed(12345)
Output = binspp(X, control, x_left, x_right, y_bottom, y_top, W_dil, z_beta, 
          z_alpha, z_omega)
\end{example}

Text output, providing the parameter estimates (medians of the estimated posterior distributions) together with the corresponding 2.5\% and 97.5\% quantiles, is obtained by the command

\begin{example}
print.outputs(Output)
\end{example}

\begin{figure}
    \centering
    \includegraphics[width=0.8\textwidth]{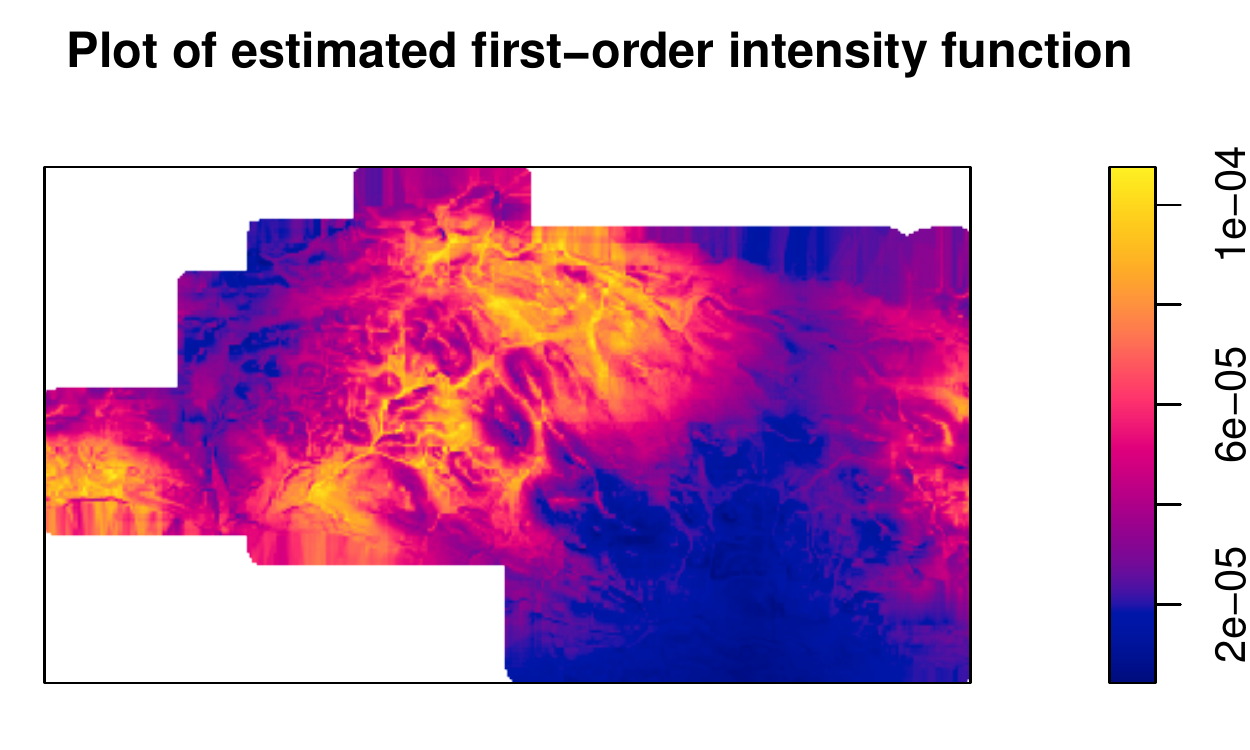}
    \includegraphics[width=0.8\textwidth]{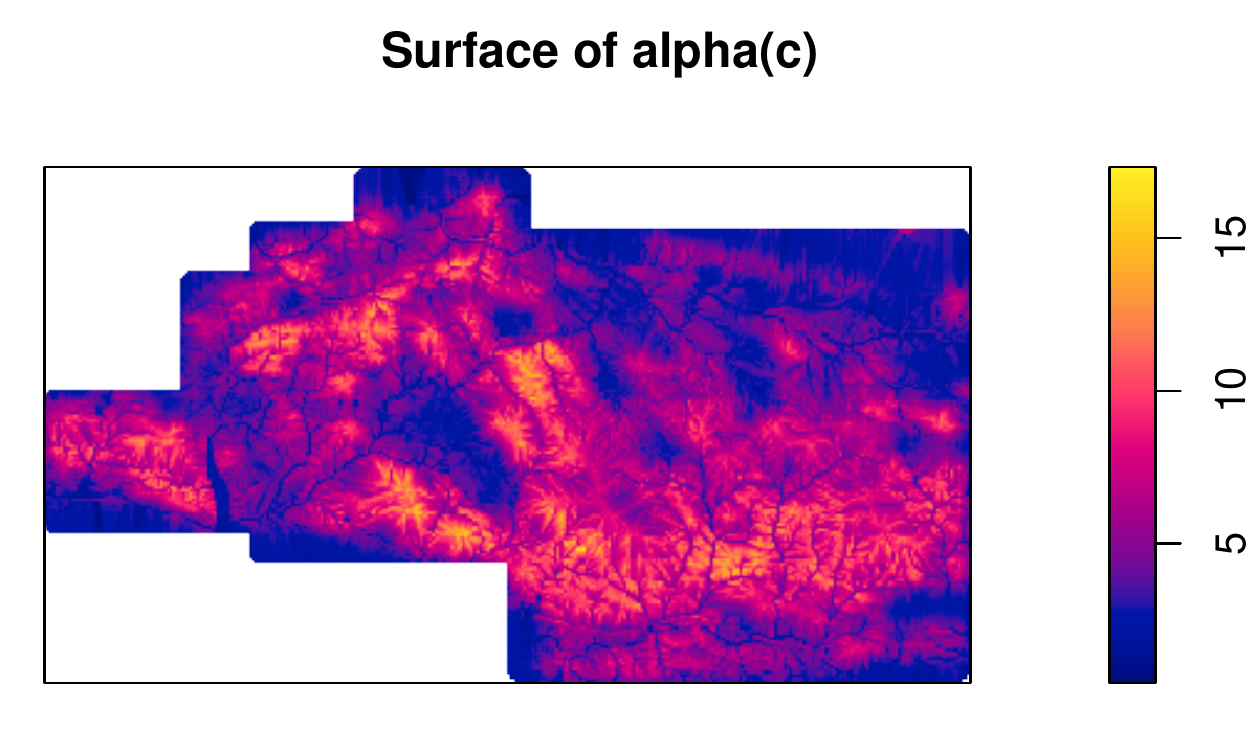}
    \includegraphics[width=0.8\textwidth]{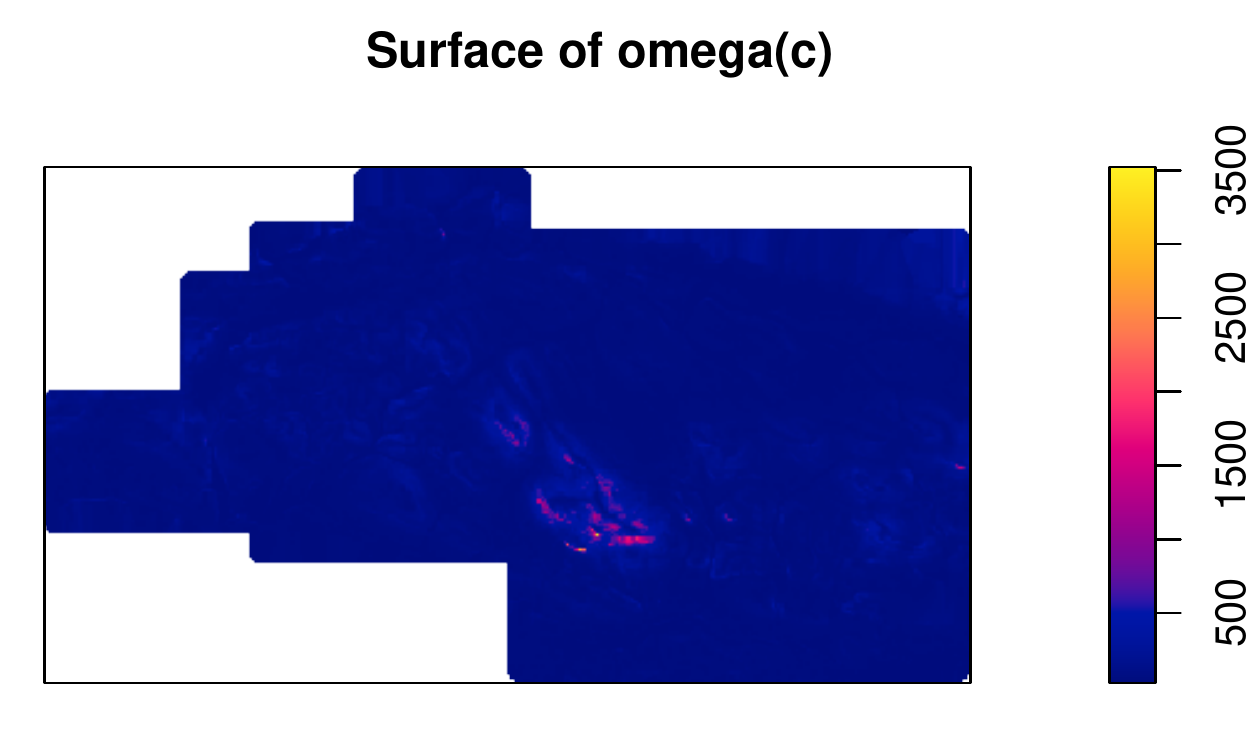}
    \caption{Plots of the estimated first-order intensity function (top), the estimated function $\alpha(c)$ describing the mean number of points in a cluster (middle) and the estimated function $\omega(c)$ describing the spread of the clusters (bottom).}
    \label{fig:binspp_outputs1}
\end{figure}

\begin{figure}
    \centering
    \includegraphics[width=0.49\textwidth]{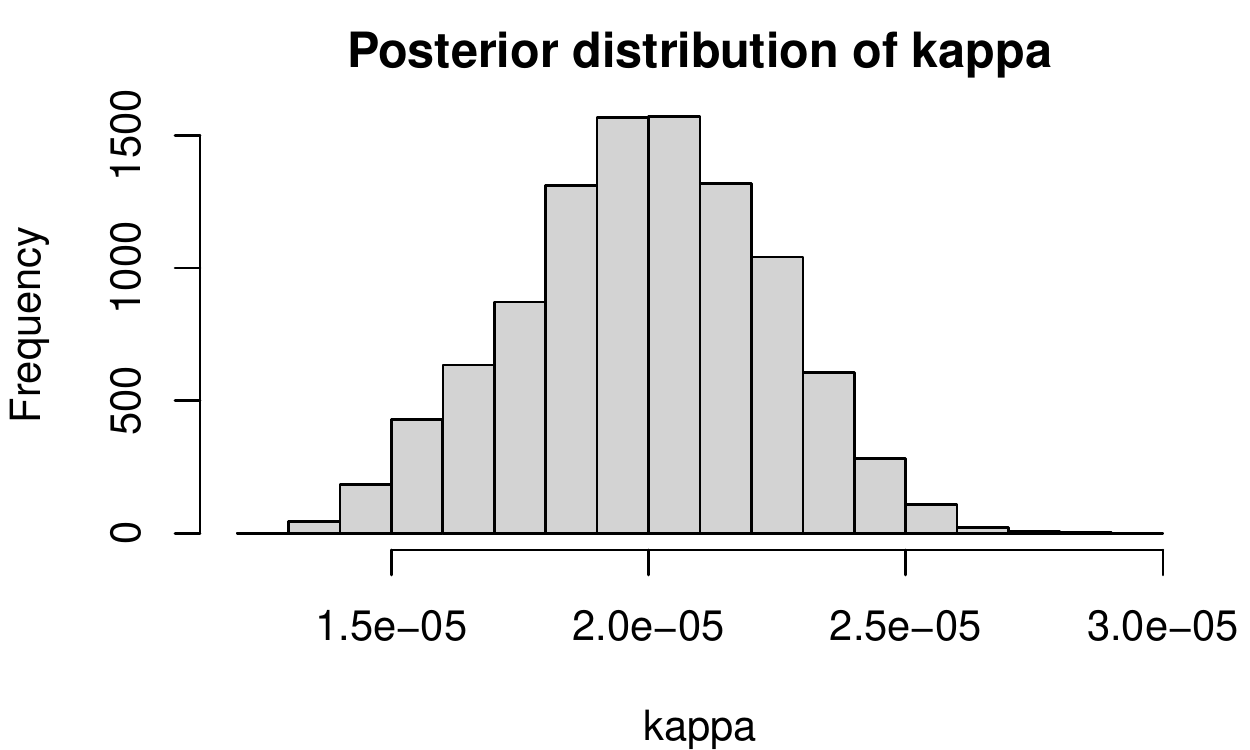}
    \includegraphics[width=0.49\textwidth]{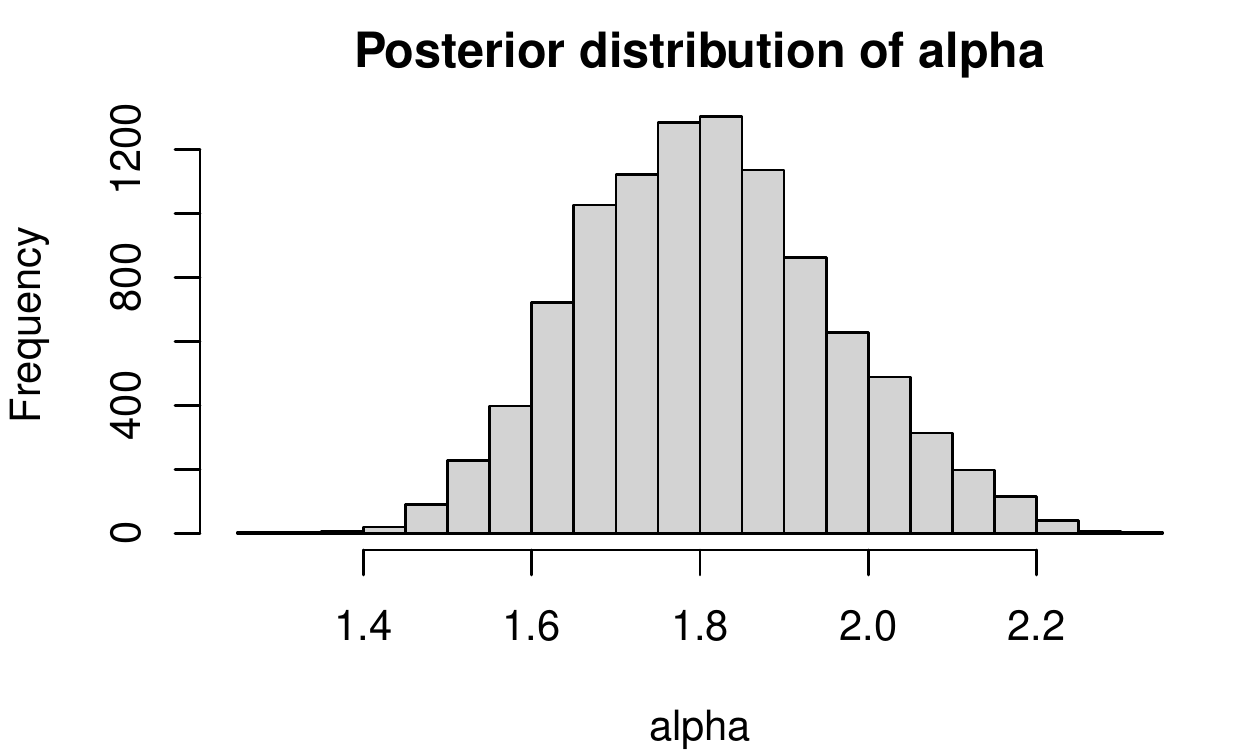}
    \includegraphics[width=0.49\textwidth]{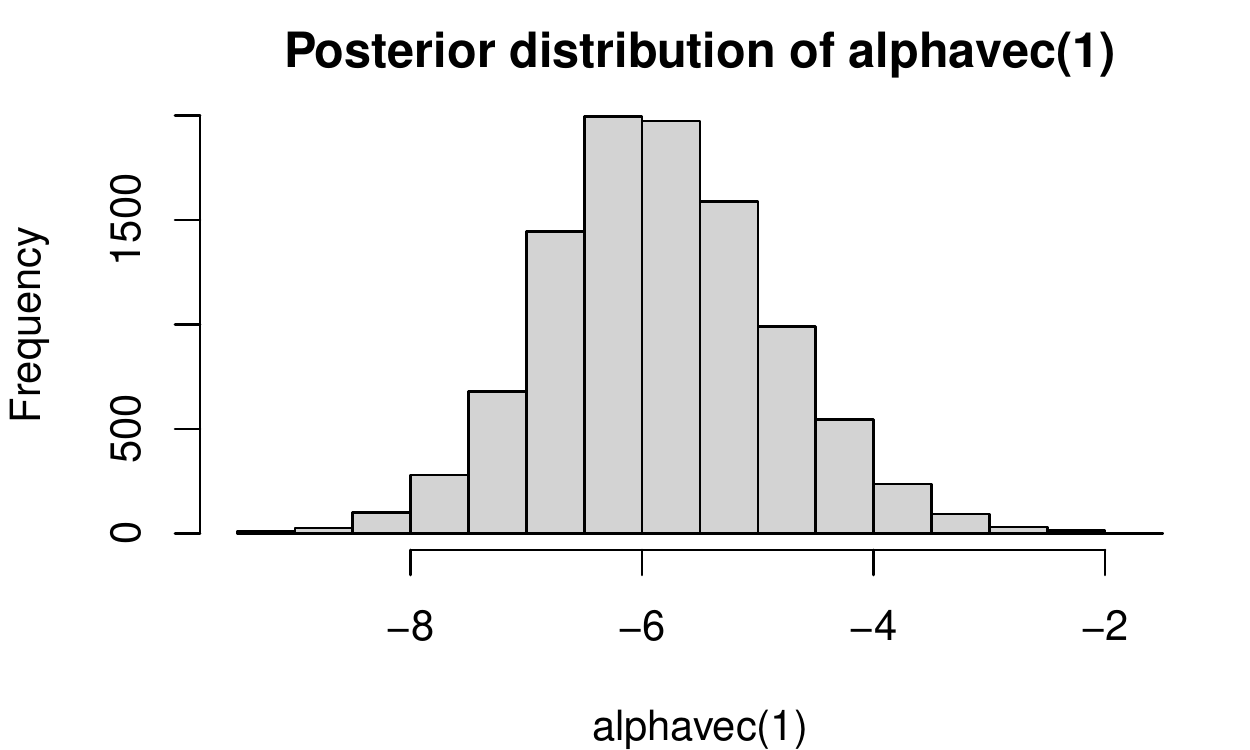}
    \includegraphics[width=0.49\textwidth]{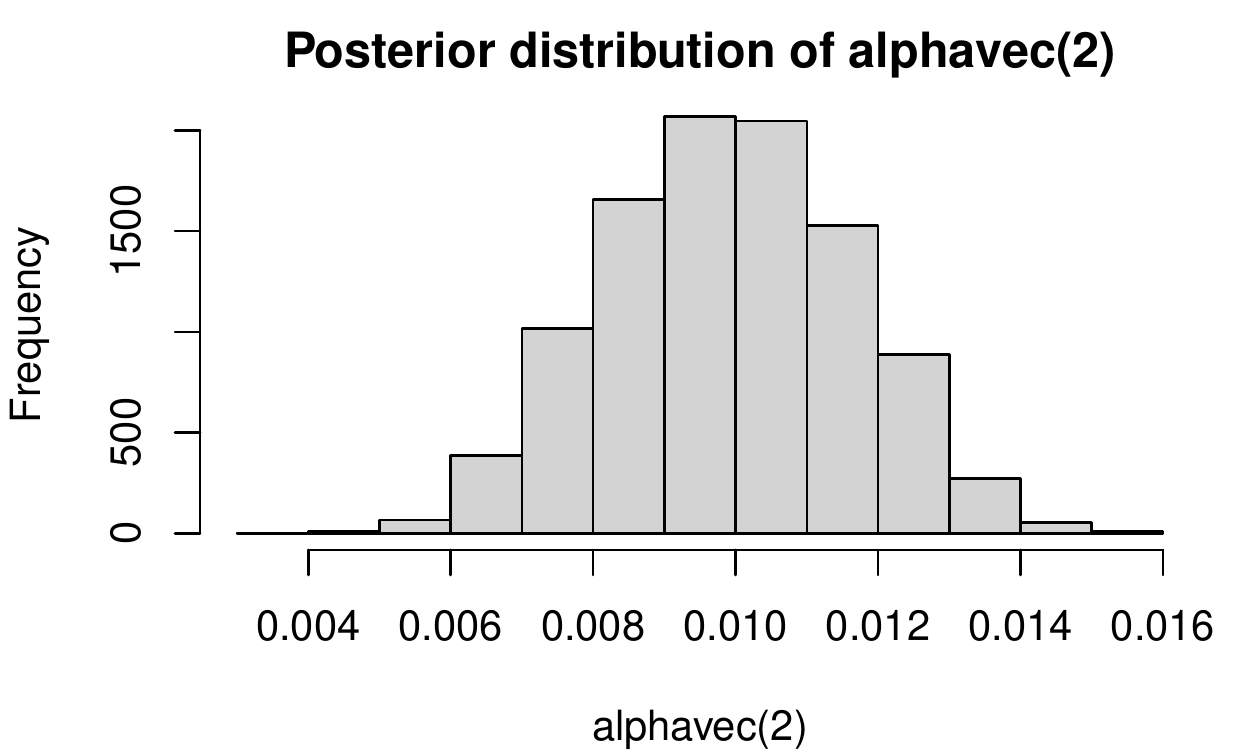}
    \includegraphics[width=0.49\textwidth]{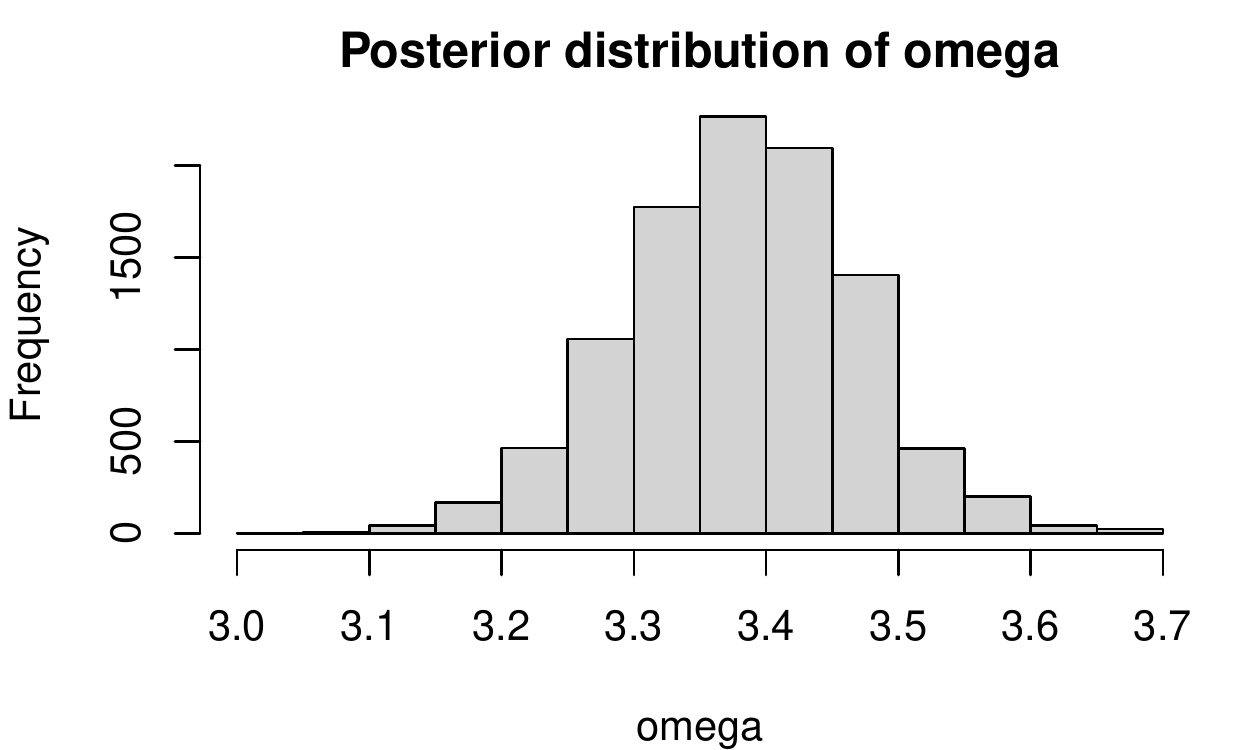}
    \includegraphics[width=0.49\textwidth]{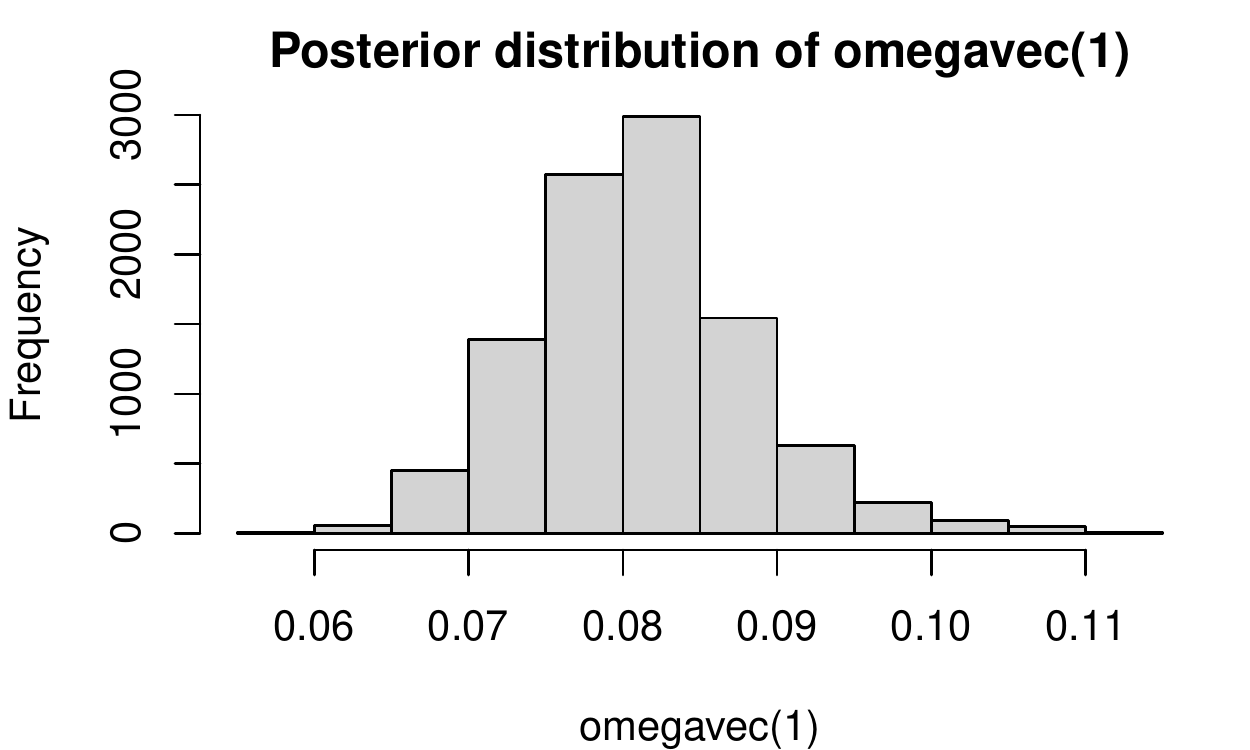}
    \includegraphics[width=0.49\textwidth]{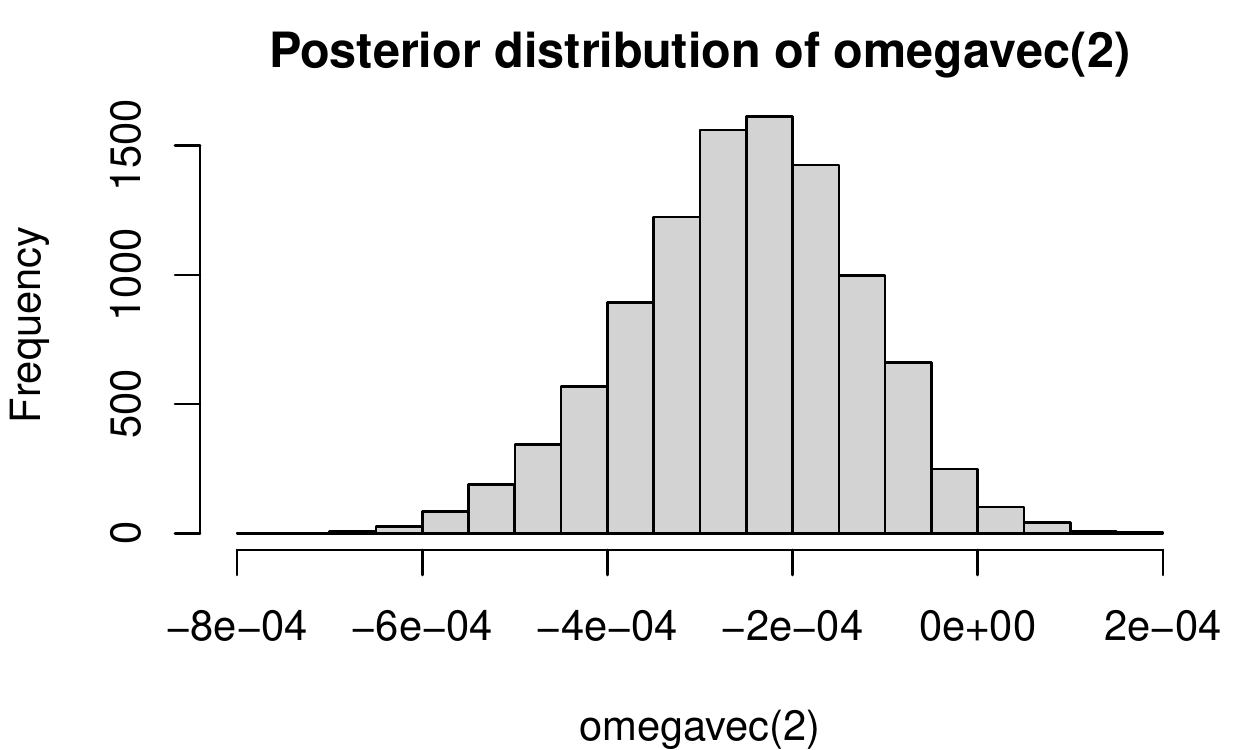}
    \includegraphics[width=0.49\textwidth]{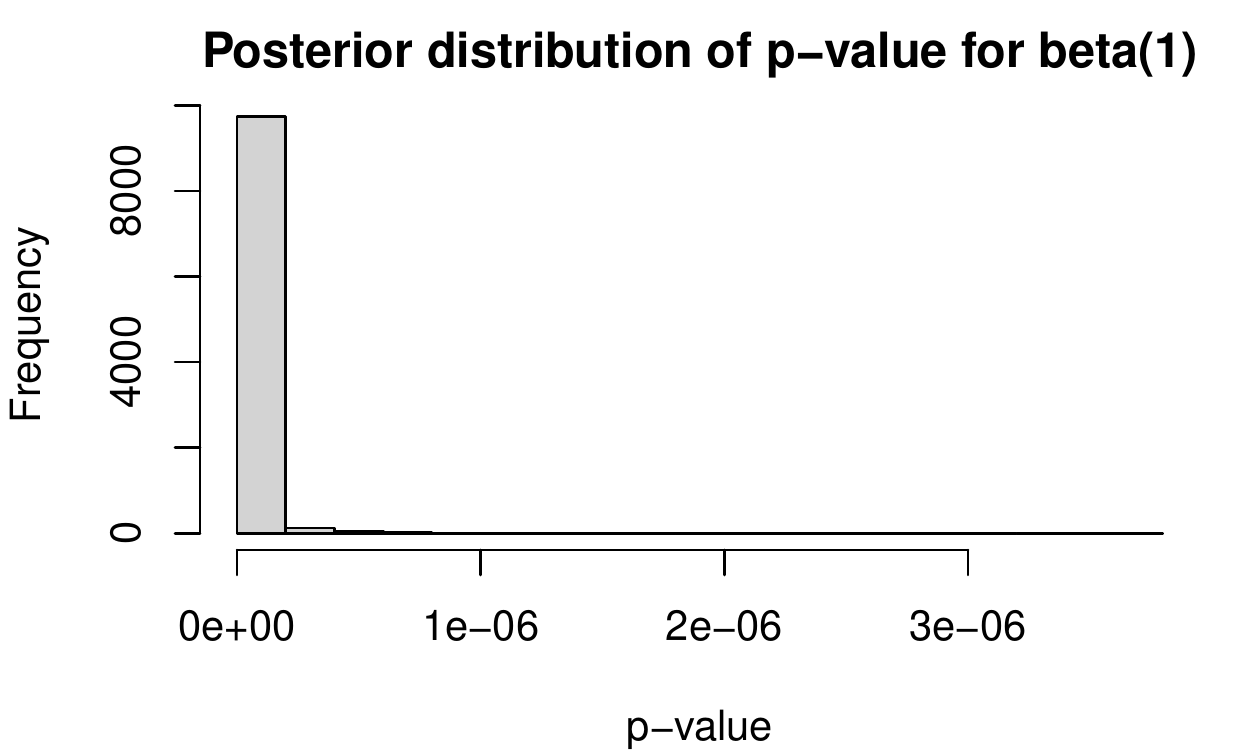}
    \includegraphics[width=0.49\textwidth]{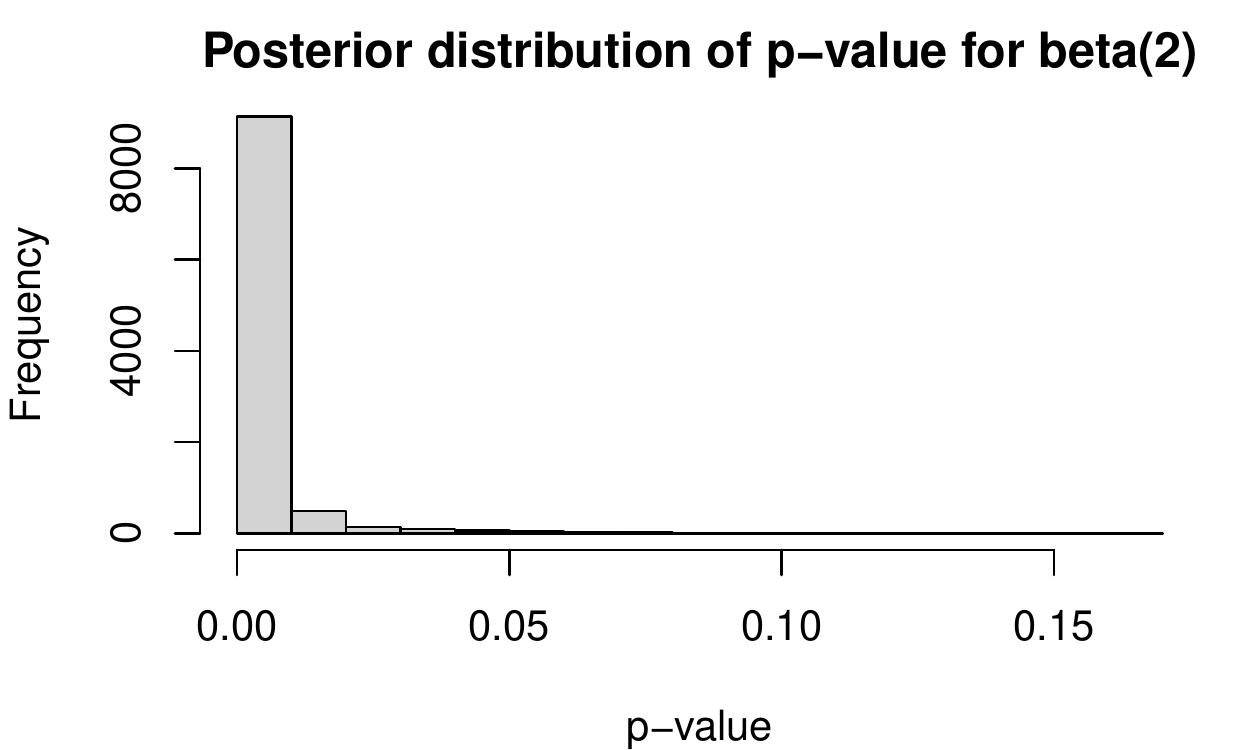}
    \caption{Histograms for estimated posterior distributions of various model parameters. Note that the parameter \texttt{alpha} in the histogram above corresponds to the parameters $\beta_0^\alpha$, the parameter \texttt{omega} corresponds to $\beta_0^\omega$ and the parameters \texttt{alphavec(i)} corresponds to the parameters $\beta_i^\alpha$ from Section~\ref{models}, \texttt{omegavec(i)} corresponds to $\beta_i^\omega$ and \texttt{beta(i)} corresponds to $\beta_i$. }
    \label{fig:binspp_outputs2}
\end{figure}

\begin{figure}
    \centering
    \includegraphics[width=0.95\textwidth]{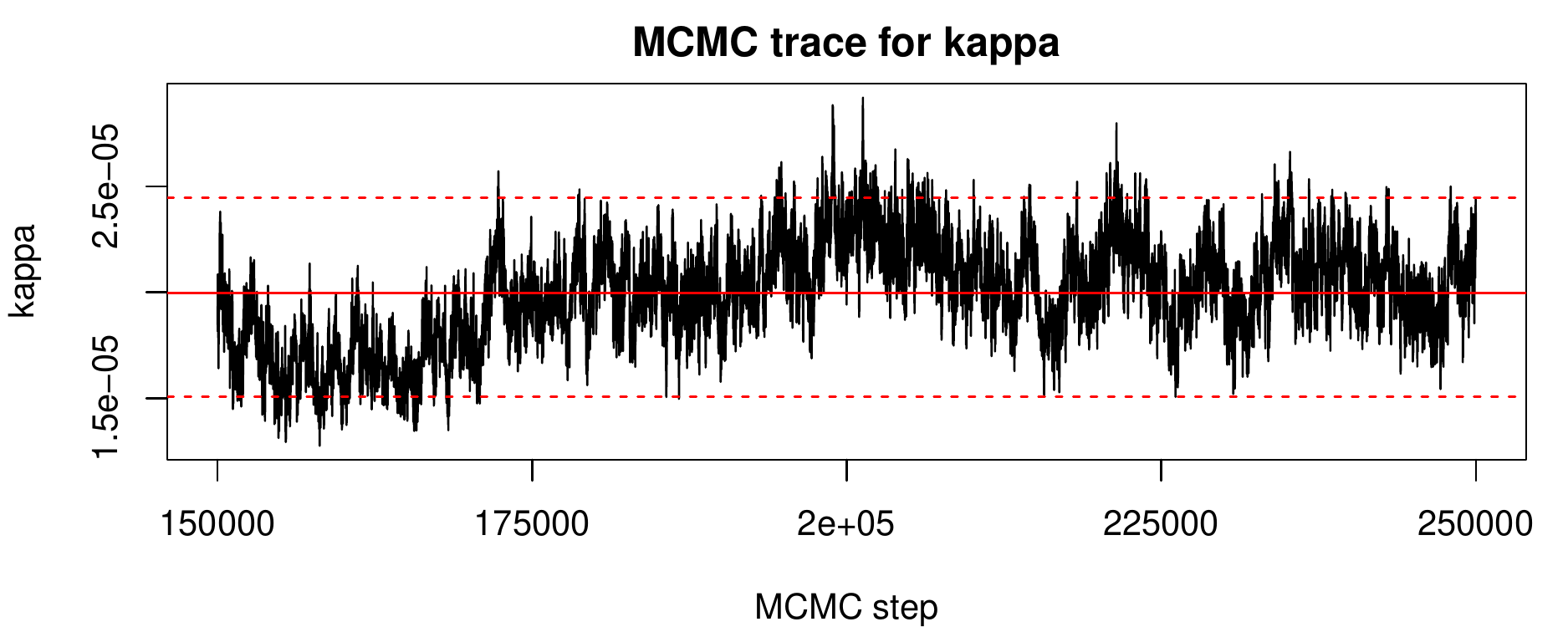}
    \includegraphics[width=0.95\textwidth]{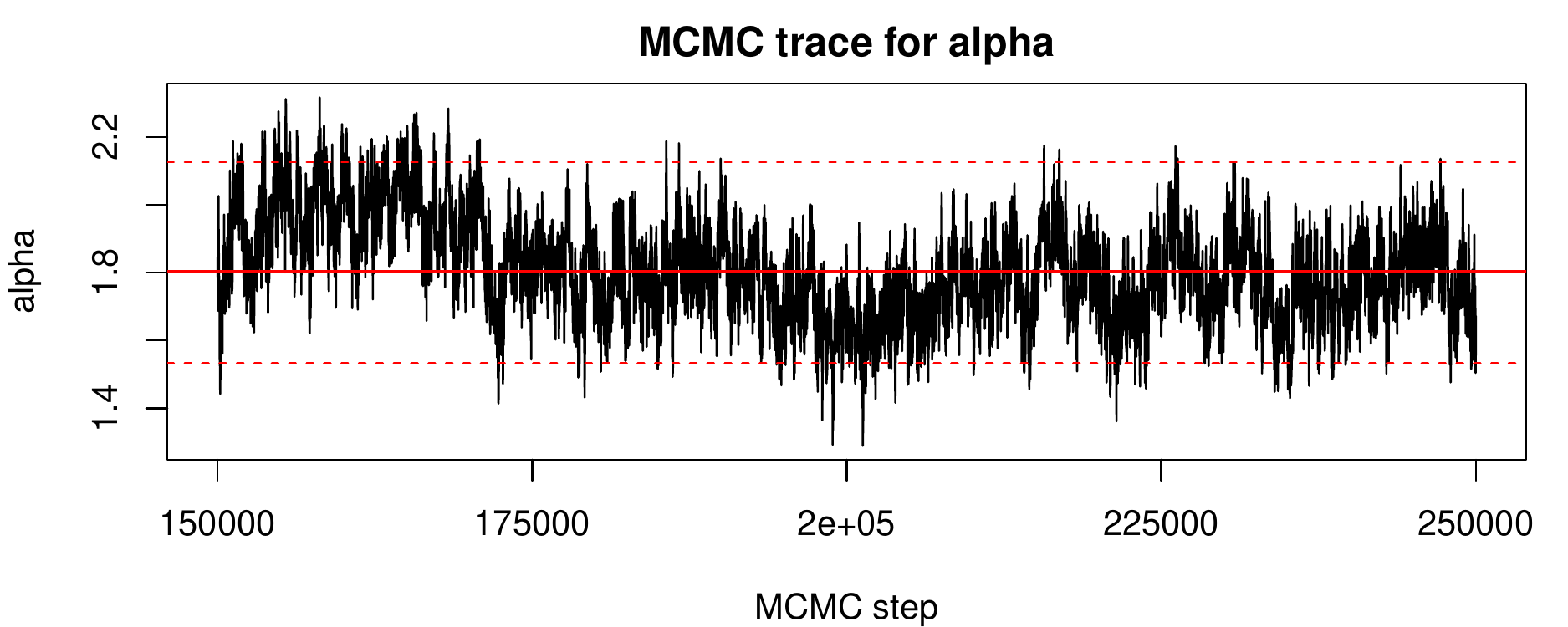}
    \includegraphics[width=0.95\textwidth]{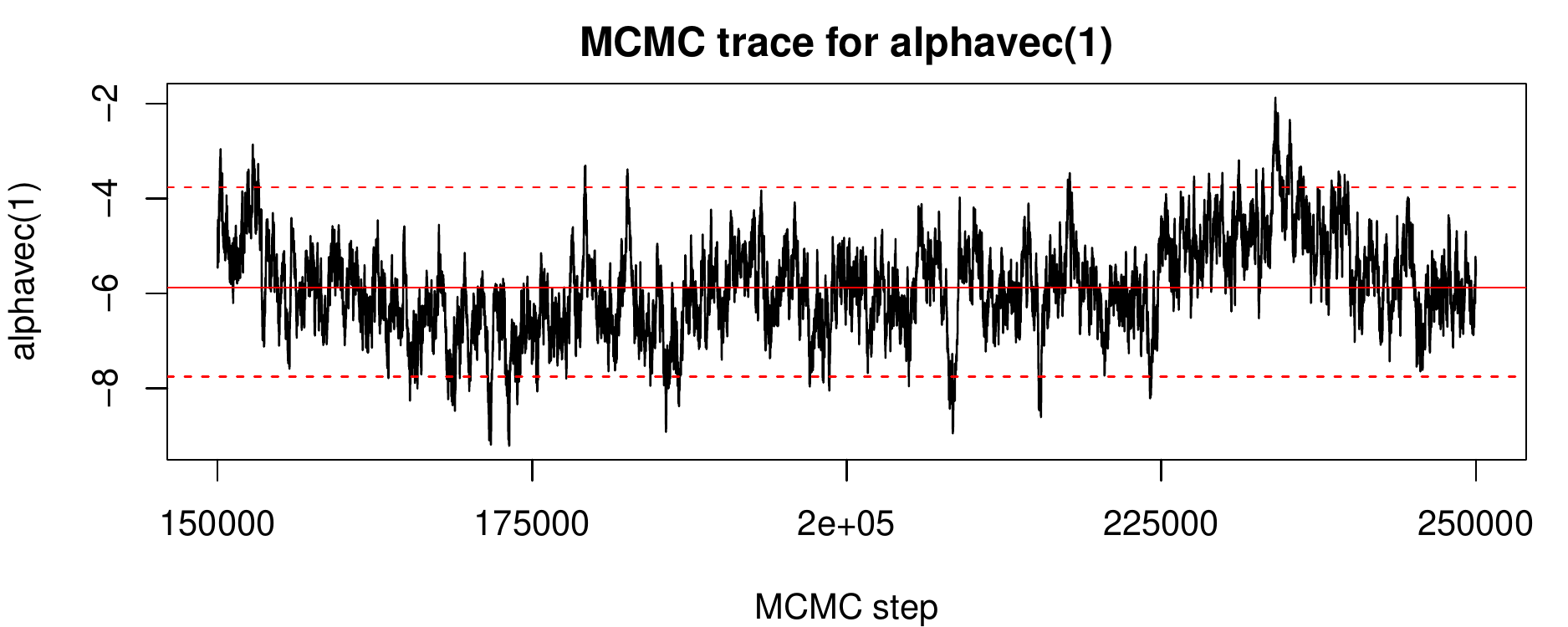}
    \includegraphics[width=0.95\textwidth]{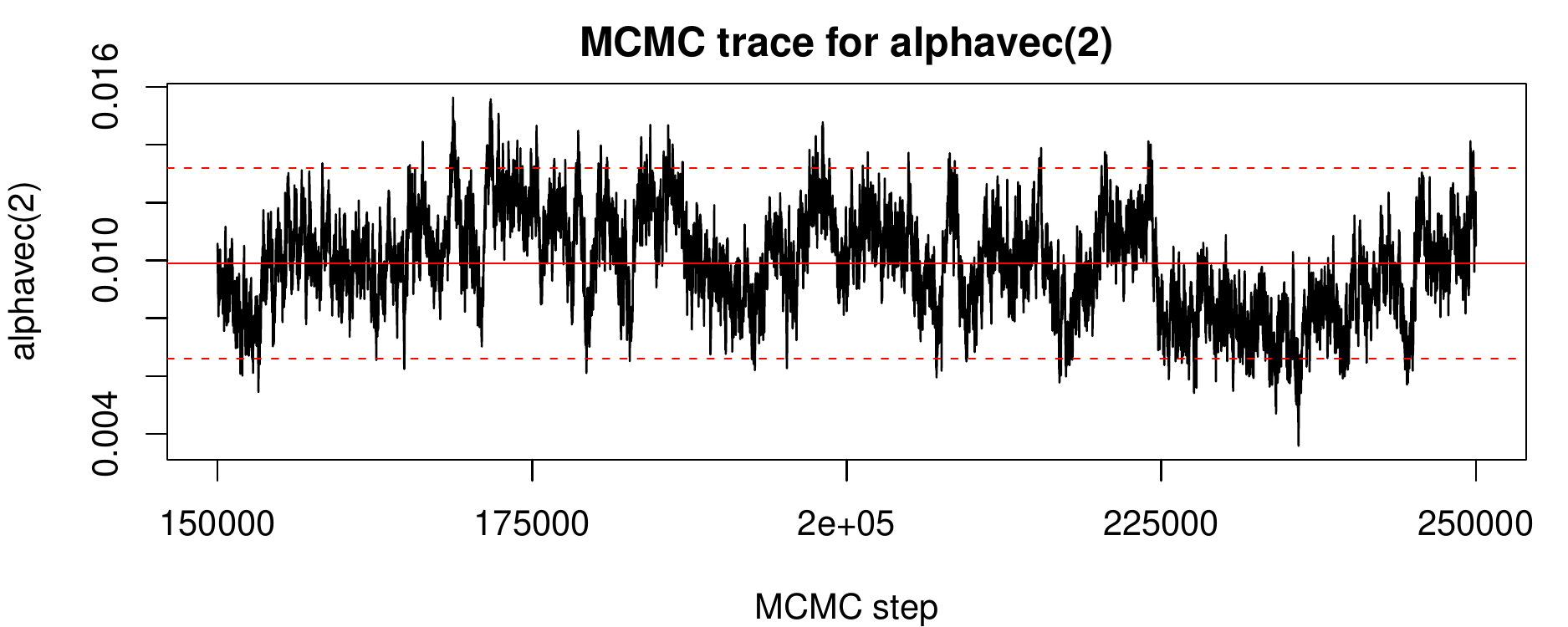}
    \caption{Traceplots for various quantities describing the state of the MCMC algorithm.}
    \label{fig:binspp_outputs3}
\end{figure}

\begin{figure}
    \centering
    \includegraphics[width=0.95\textwidth]{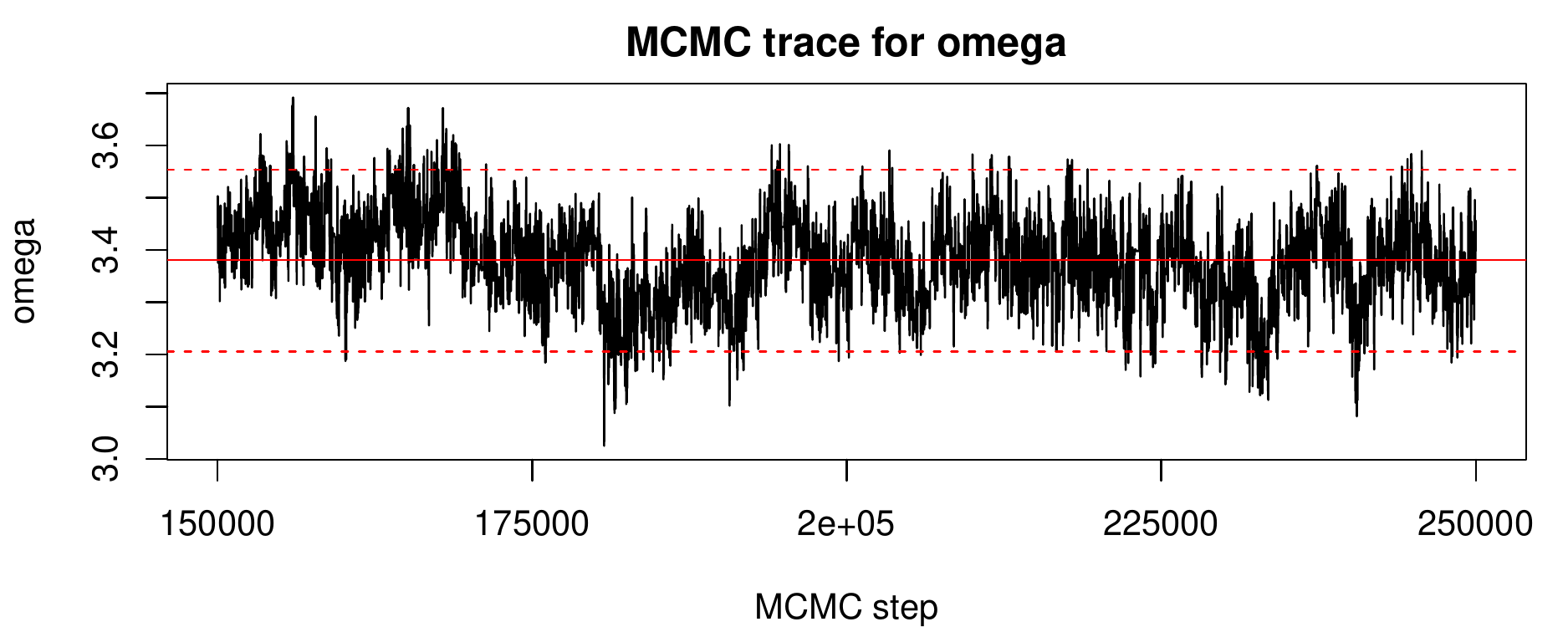}
    \includegraphics[width=0.95\textwidth]{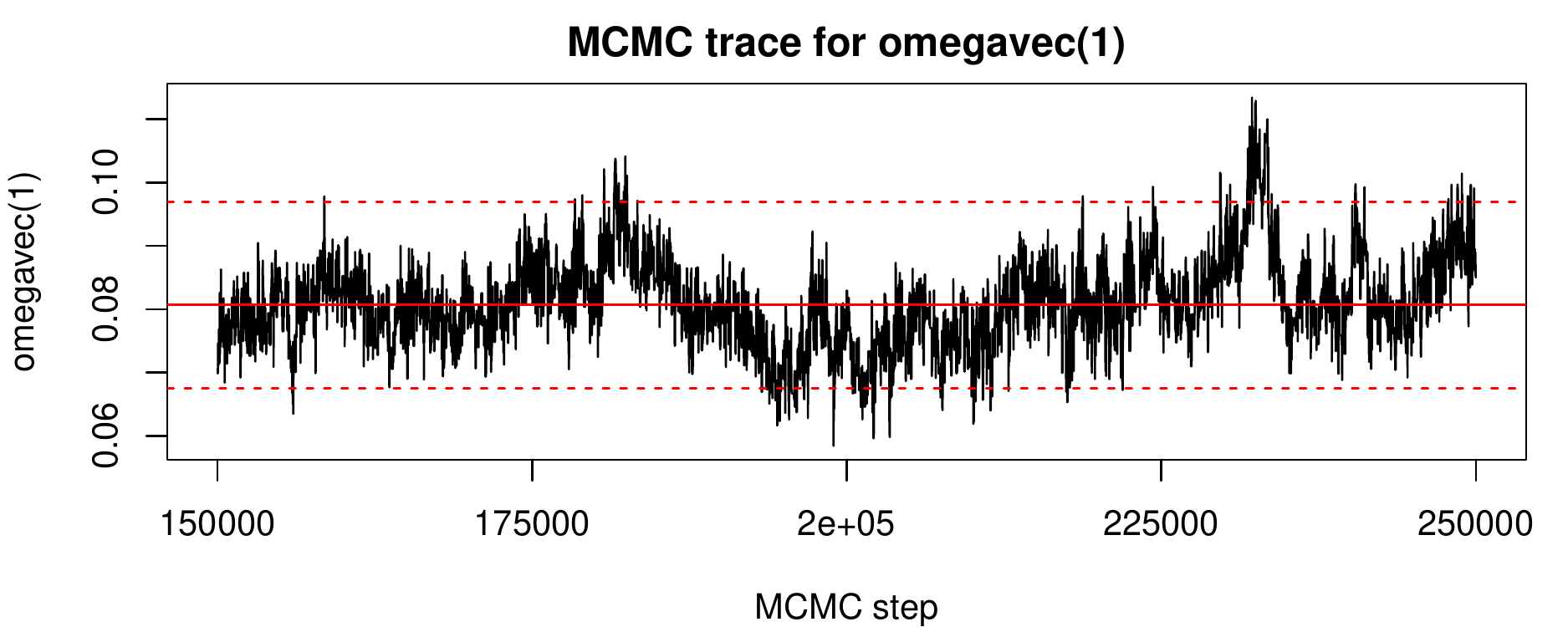}
    \includegraphics[width=0.95\textwidth]{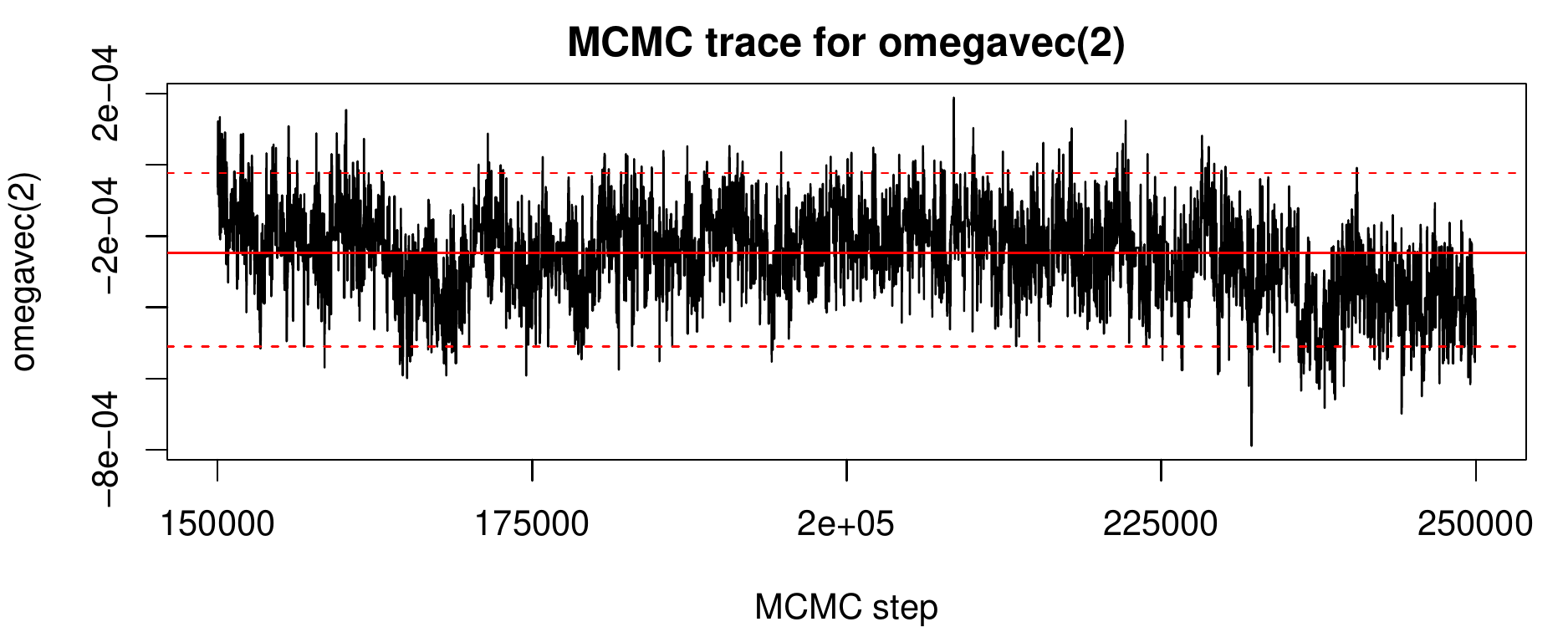}
    \includegraphics[width=0.95\textwidth]{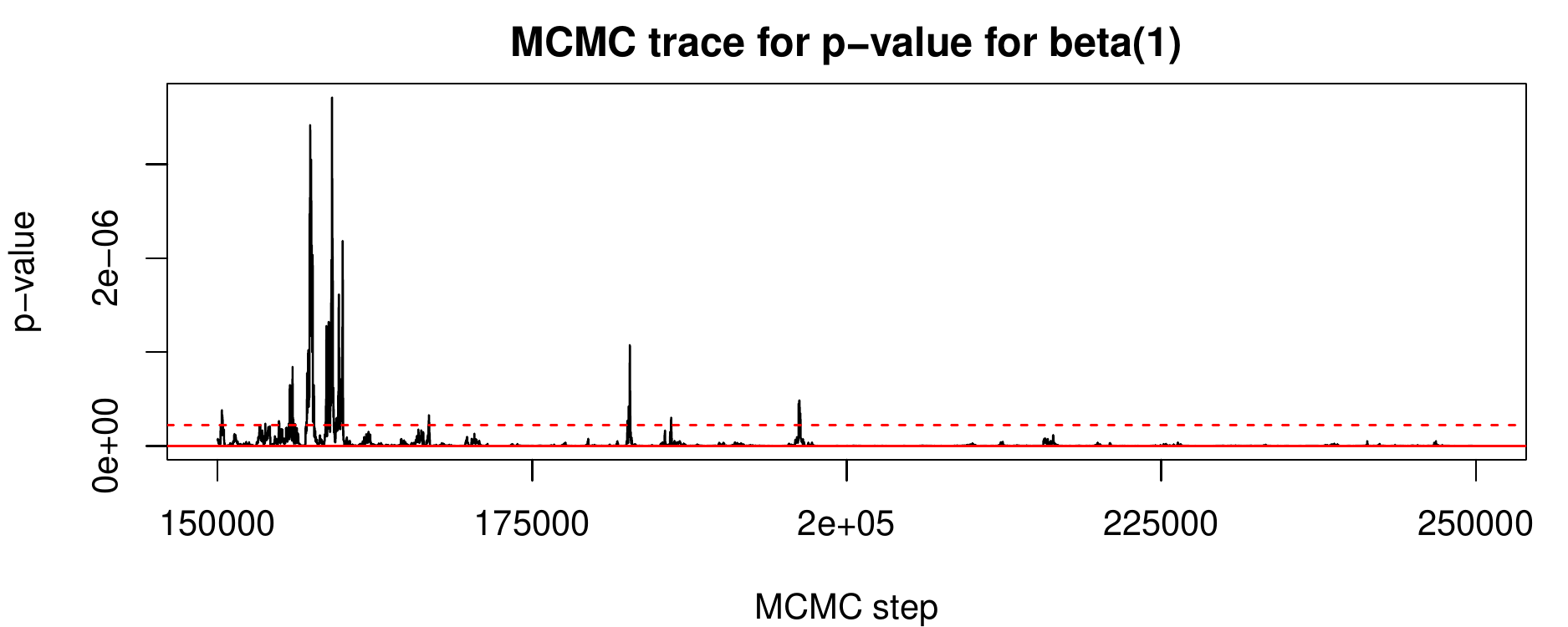}
    \caption{Traceplots for various quantities describing the state of the MCMC algorithm.}
    \label{fig:binspp_outputs4}
\end{figure}

\begin{figure}
    \centering
    \includegraphics[width=0.95\textwidth]{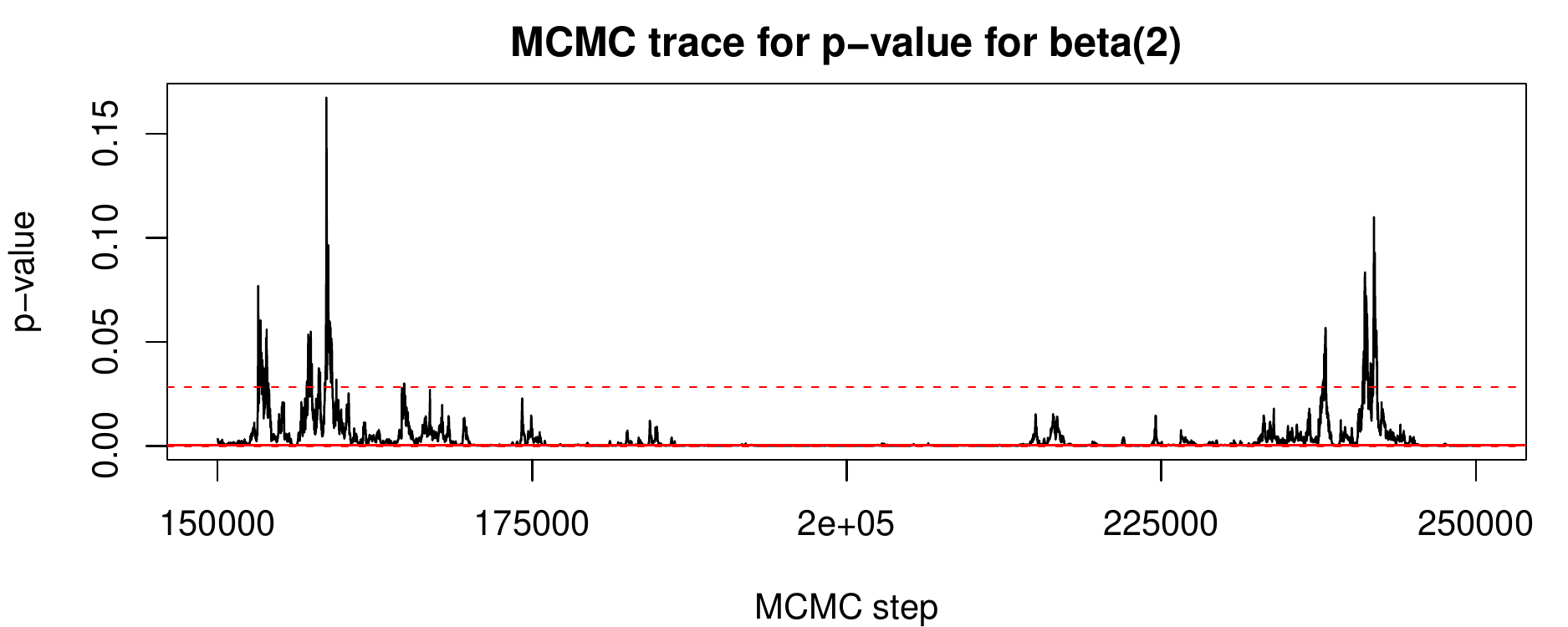}
    \includegraphics[width=0.95\textwidth]{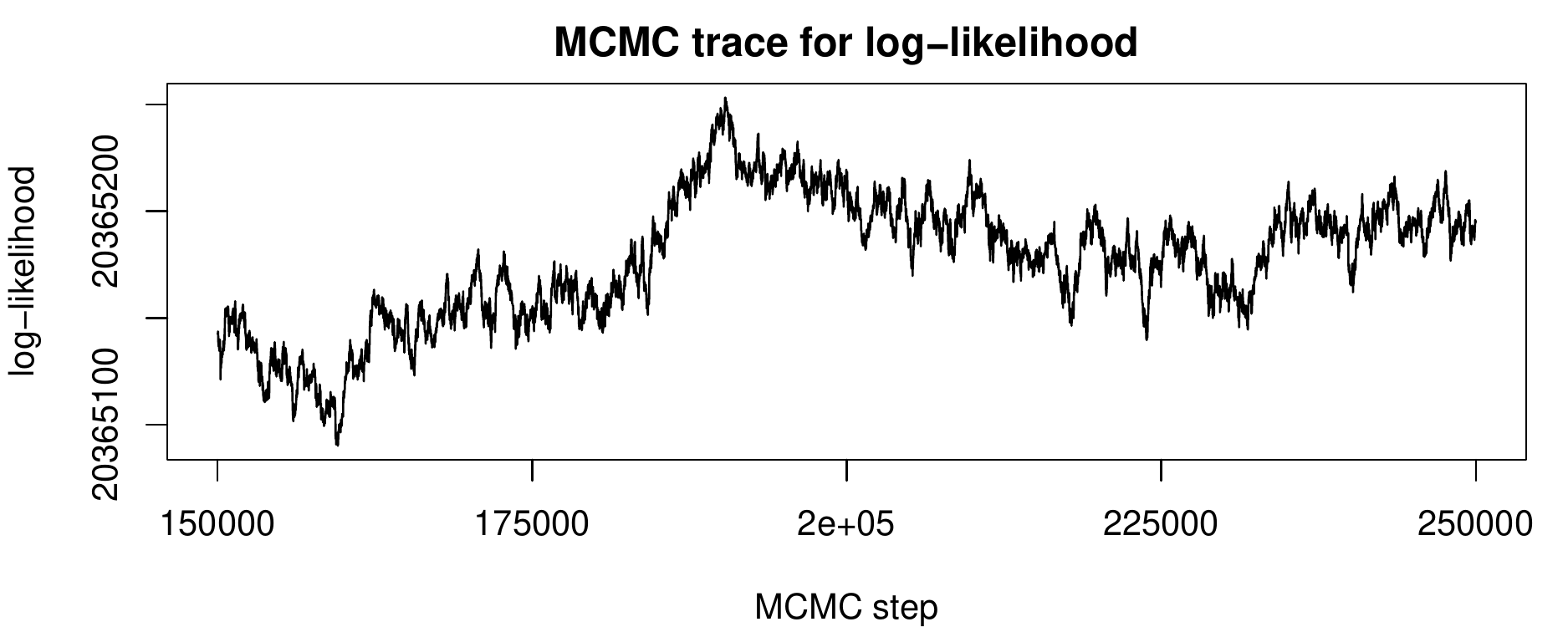}
    \includegraphics[width=0.95\textwidth]{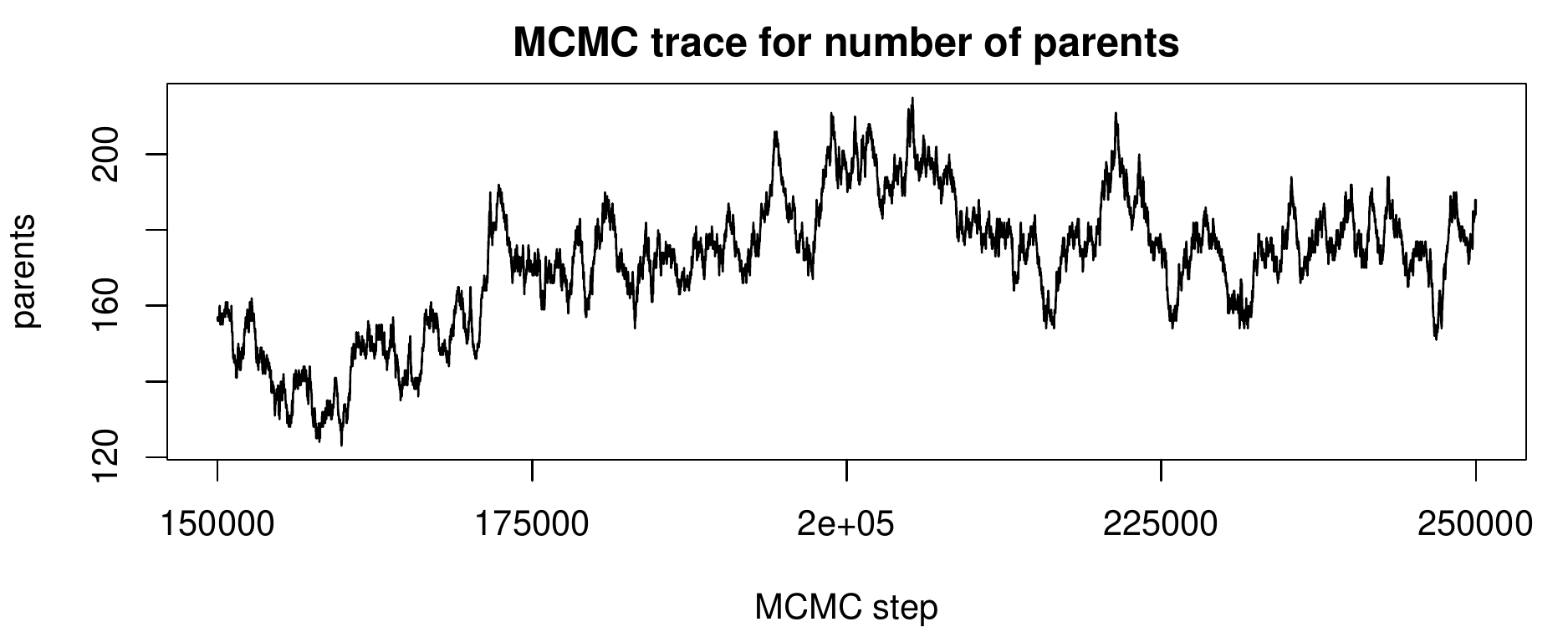}
    \includegraphics[width=0.95\textwidth]{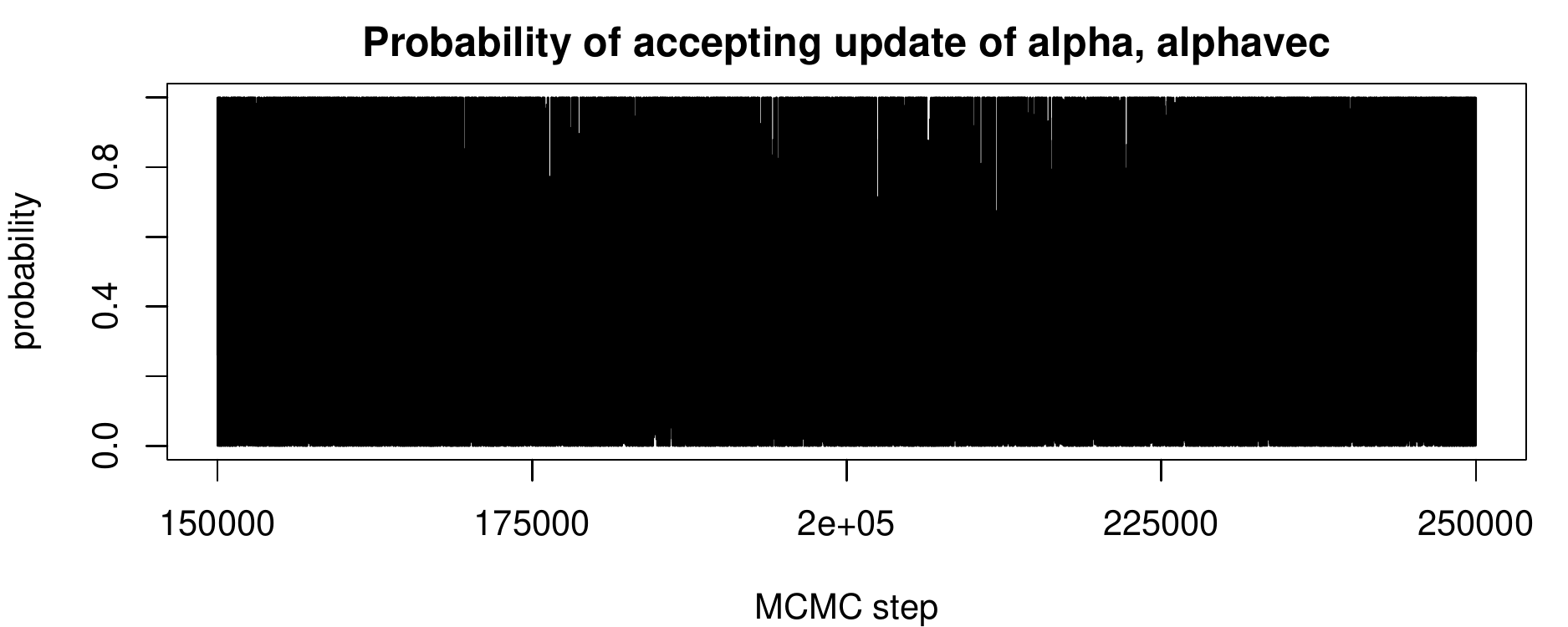}
    \caption{Traceplots for various quantities describing the state of the MCMC algorithm.}
    \label{fig:binspp_outputs5}
\end{figure}

\begin{figure}
    \centering
    \includegraphics[width=0.95\textwidth]{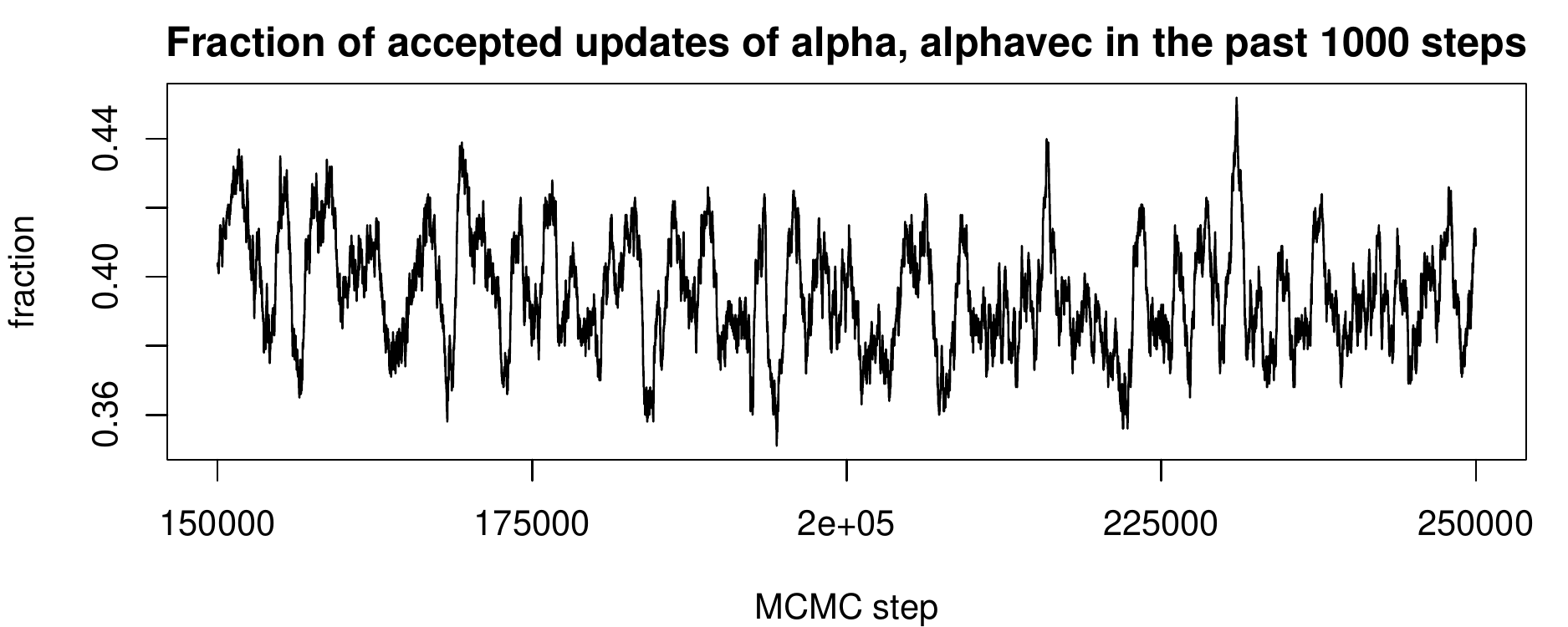}
    \includegraphics[width=0.95\textwidth]{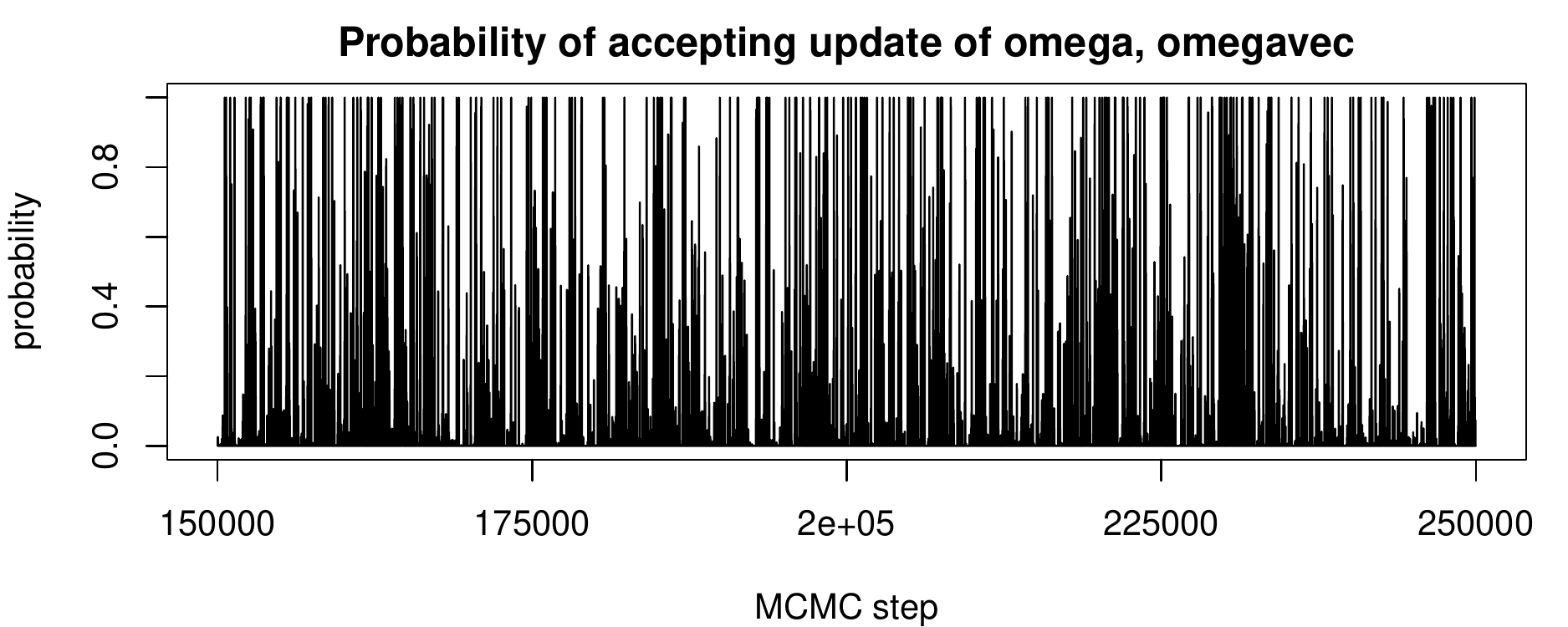}
    \includegraphics[width=0.95\textwidth]{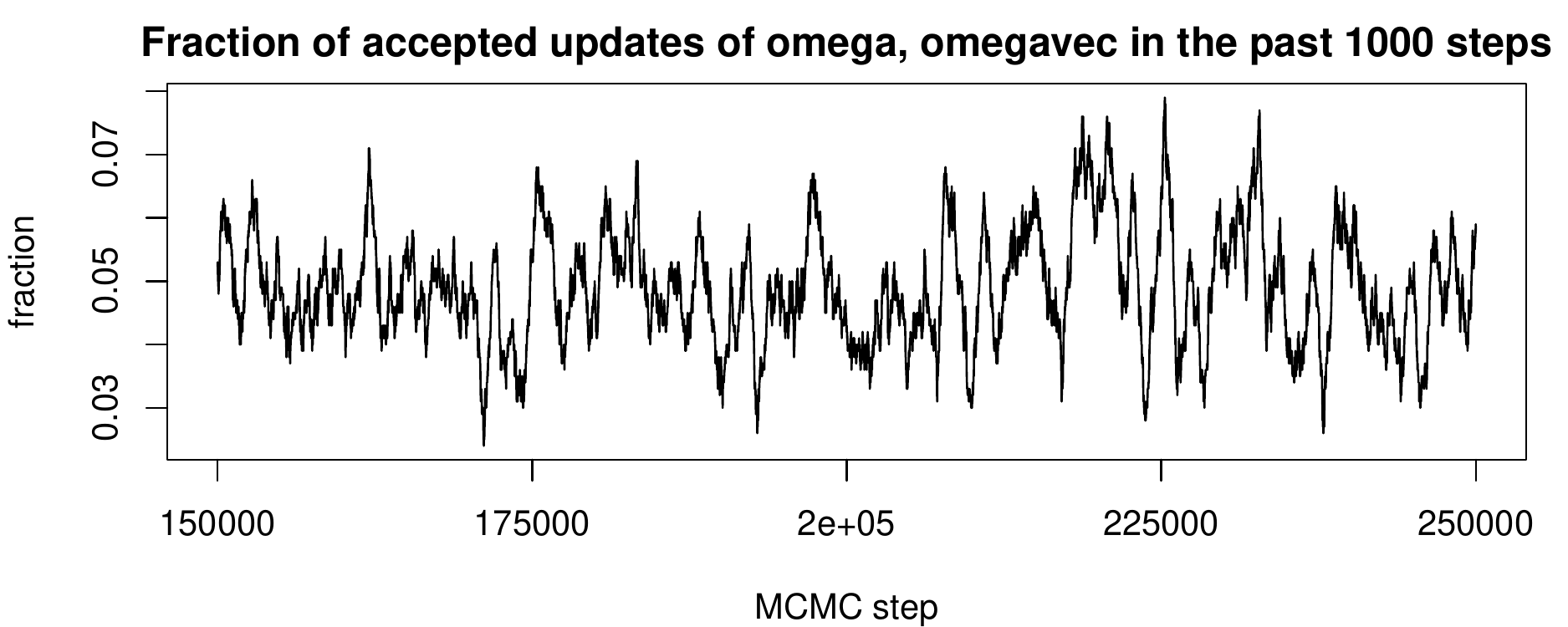}
    \includegraphics[width=0.95\textwidth]{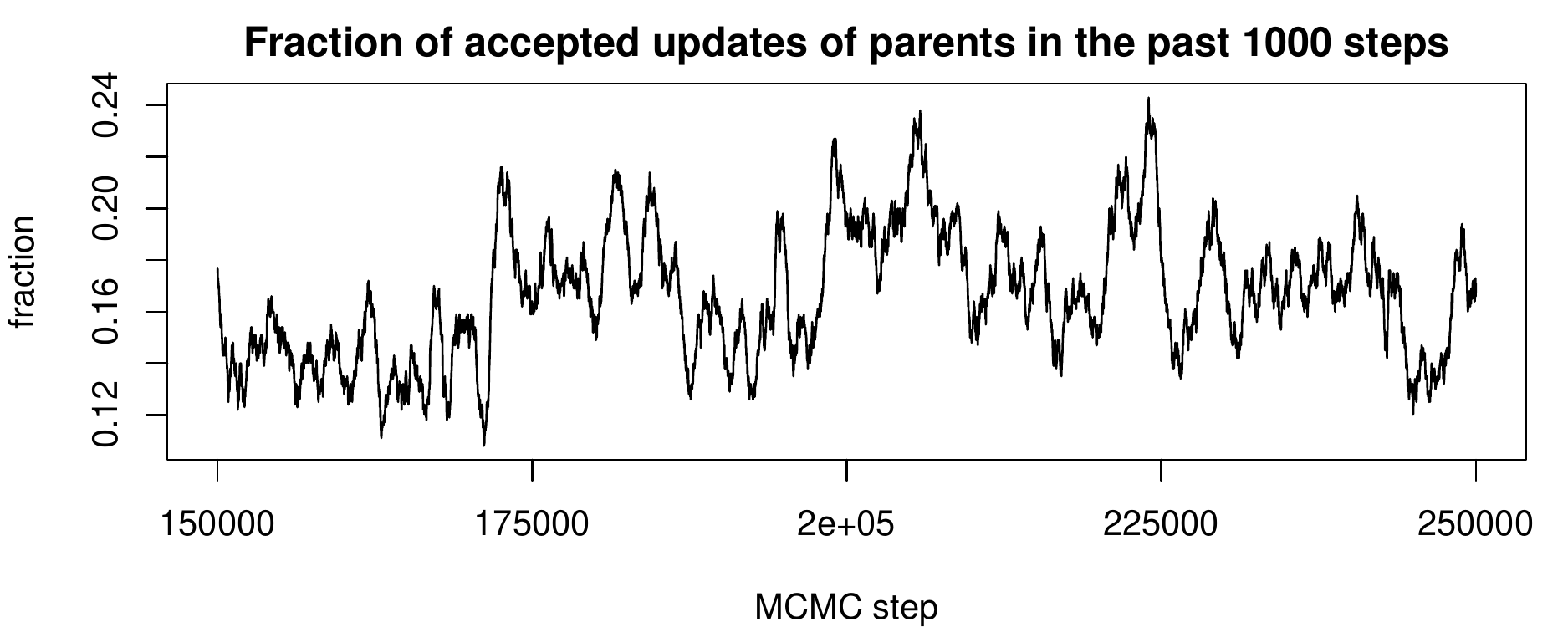}
    \caption{Traceplots for various quantities describing the state of the MCMC algorithm.}
    \label{fig:binspp_outputs6}
\end{figure}

Graphical output, given in Figures~\ref{fig:binspp_outputs1} to \ref{fig:binspp_outputs6}, is provided by the command

\begin{example}
plot.outputs(Output)
\end{example}

First, estimated surfaces of the first-order intensity function, the $\alpha(c)$ function, describing the mean number of points in a cluster, and the $\omega(c)$ function, describing the spread of the clusters, are plotted, as illustrated in Figure~\ref{fig:binspp_outputs1}.

Then, histograms describing the estimated posterior distribution of the model parameters are plotted, see Figure~\ref{fig:binspp_outputs2}. 

The histograms for the first order parameters are not plotted, because the posterior distribution of these  $\beta_1$ and $\beta_2$ are computed in the first under the assumption of Poissonity. But these parameter are bound with inhomogeneity of cluster centers, i.e. its significance should be computed from cluster centers only. In order to deal with this issue, we record the centers in every step of MCMC and also we record the significance of the covariates from the list \texttt{z\char`_beta} with respect to the population of cluster centers in every step of MCMC. Posterior histograms of the corresponding p-values are plotted instead of the posterior histograms for these parameters, see the bottom part of Figure~\ref{fig:binspp_outputs2}. This provides more precise inference about the significance of the covariates than the outcomes of the first step of estimation, where the \texttt{ppm} function from the \pkg{spatstat} package is used and where the precise locations of the (observed) offsprings may confound the significance of the covariates with respect to the (unobserved) parent points.

Finally, traceplots for various quantities describing the state of the chain are plotted, including 1) the model parameters, 2) the p-values discussed in the previous paragraph, 3) the value of the log-likelihood itself, 4) the number of parent points, 5) acceptance probabilities for proposed updates of parameters influencing $\alpha(c)$ or $\omega(c)$ (not very informative for long runs but useful when tuning the algorithm for a new dataset with short runs), and 6) fractions of accepted updates in the past 1000 steps of the algorithm (much more informative than plots of the acceptance probabilities). See Figures~\ref{fig:binspp_outputs3} to \ref{fig:binspp_outputs6} for illustration. The traceplots for model parameters describe also the estimated median of the posterior distribution (given by the solid red line) together with the empirical 2.5\% and 97.5\% quantiles of the posterior distribution (red dashed lines).

\subsection{Generalised Neyman-Scott}\label{subsec:generalised}

In this subsection we specify the implementation of the Bayesian MCMC algorithm for estimation of the homogeneous generalised Thomas point process (GTPP) which was described in \cite{AM2020}. In this subsection we assume that the point pattern \texttt{X} is an object of the format \texttt{ppp} from the \pkg{spatstat} package and that it is observed in a nrectangular observation window $W$. The notation and priors are slightly different than for the inhomogeneous models, due to the different notations used in the {corresponding} original papers. 

The GTPP can be simulated using

\begin{example}
X=rgtp(kappa, omega, lambda, theta, win = W)
\end{example}

Here \texttt{kappa} corresponds to the intensity of centers, \texttt{omega} to the standard deviation of the radially symmetric Gaussian distribution determining the spread of offsprings, and  \texttt{lambda} and \texttt{theta} correspond to the parameters of the GPD {governing the} cluster size.

The priors used in the estimation are lognormal for all parameters, except \texttt{lambda} which has a uniform prior. The hyperparametres and control parameters must be specified in the estimation function \texttt{estgtp} itself. The function \texttt{estgtpr} allows for plotting of all results and outputs of the MCMC chain. 

The posterior distribution of the parameter \texttt{lambda} can be used to determine the over- or under-dispersion. {Specifically, if} 0 is an element of the 95\% posterior interval for \texttt{lambda}, then the {Poisson assumption} cannot be rejected. See the example code where result summaries all posterior medians and 95\% quantiles of the model parameters.  

\begin{example}
#Prior for parameter kappa
a_kappa = 4
b_kappa = 1
x <- seq(0, 100, length = 100)
hx <- dlnorm(x, a_kappa, b_kappa)
plot(x, hx, type = "l", lty = 1, xlab = "x value",
     ylab = "Density", main = "Prior")

#Prior for parameter omega
a_omega = -3
b_omega = 1
x <- seq(0, 1, length = 100)
hx <- dlnorm(x, a_omega, b_omega)
plot(x, hx, type = "l", lty = 1, xlab = "x value",
     ylab = "Density", main = "Prior")

#Prior for parameter lambda
l_lambda = -1
u_lambda = 0.99
x <- seq(-1, 1, length = 100)

hx <- dunif(x, l_lambda, u_lambda)
plot(x, hx, type = "l", lty = 1, xlab = "x value",
     ylab = "Density", main = "Prior")

#Prior for parameter theta
a_theta = 4
b_theta = 1
x <- seq(0, 100, length = 100)
hx <- dlnorm(x, a_theta, b_theta)
plot(x, hx, type = "l", lty = 1, xlab = "x value",
     ylab = "Density", main = "Prior")

#estimation procedure with default values for standard deviations of updates for the parameters
est = estgtp(X$X,
             skappa = exp(a_kappa + ((b_kappa ^ 2) / 2)) / 100, 
             somega = exp(a_omega + ((b_omega ^ 2) / 2)) / 100, dlambda = 0.01,
             stheta = exp(a_theta + ((b_theta ^ 2) / 2)) / 100, smove = 0.1,
             a_kappa = a_kappa, b_kappa = b_kappa,
             a_omega = a_omega, b_omega = b_omega,
             l_lambda = l_lambda, u_lambda = u_lambda,
             a_theta = a_theta, b_theta = b_theta,
             iter = 1000, plot.step = 1000, save.step = 1e9,
             filename = "")

#plotting the results and estimationg of the parameters from refined MCMC chain
discard = 100
step = 10

result = estgtpr(est, discard, step)
result

\end{example}

\section{Summary}\label{sec:discussion}

This paper presents a model and an implementation of the Bayesian MCMC algorithm for estimating parameters of inhomogeneous Neyman-Scott point process with inhomogeneity in cluster centers, cluster spread and/or cluster sizes. The simulation studies {assessing} the performance of the presented algorithm were published in \cite{KM2016} for the homogeneous process, in \cite{MMK2014} for inhomogeneous cluster centers and in \cite{MS2017} for inhomogeneous cluster centers and cluster spread.

The presented package \pkg{binspp} allows also for identification of over- or under-dispersion of cluster sizes in a homogeneous model through the generalised Thomas point process. Since the algorithm is rather flexible, the package can be further {extended} by estimation methods for models where some cluster centers are given and fixed. It can also be {extended} into the direction of the spatio-temporal Neyman-Scott point processes. 

A user of this package should be aware of the following issues, connected with the Bayesian MCMC algorithms. First, priors should be carefully chosen, even though we provide a {sensible} default. For example, the range of the $\omega$ prior must be in coherence with the size of the observation window; similarly, the range of the priors for the regression parameters connected with the covariates must reflect the range of the covariate values. Second, when {slow} mixing is observed via low fractions of accepted updates, one might think of changing the standard deviation for proposal distributions. These standard deviations are also provided in the package by their default value. Third, if many covariates are {considered}, longer chains are needed {to get reasonably close to the equilibrium}. The equilibrium state can be determined by {the log-likelihood not increasing anymore}, {stabilized} number of centers, {stationary behaviour of the trace plots of the parameters}, or by {unimodal} histograms of posterior distributions.

\section*{Acknowledgements}
The project has been financially supported by the Grant Agency of Czech Republic (Project No. 19-04412S). The authors are very grateful to Begona Abellanas who initiated this project by asking for the implementation of the method and who provide us with the illustratory data set.

\address{Jiří Dvořák\\
  Charles univerzity, Faculty of Mathematics and Physics\\
 Sokolovská 83, 186 75 Prague \\
  Czech Republic\\
  (ORCiD: 0000-0003-3290-8518)\\
  \email{dvorak@karlin.mff.cuni.cz}}

\address{Radim Remeš\\
  University of South Bohemia, Faculty of Economics\\
  Studentská 13\\
  Czech Republic\\
  (ORCiD: 0000-0002-3306-0347)\\
  \email{inrem@ef.jcu.cz}}

\address{Ladislav Beránek\\
  University of South Bohemia, Faculty of Economics\\
  Studentská 13\\
  Czech Republic\\
  (ORCiD: 0000-0001-5004-0164)\\
  \email{beranek@ef.jcu.cz}}

\address{Tomáš Mrkvička\\
  University of South Bohemia, Faculty of Economics\\
  Studentská 13\\
  Czech Republic\\
  (ORCiD: 0000-0003-1613-2780)\\
  \email{mrkvicka.toma@gmail.com}}

\end{article}

\end{document}